\documentclass[preprint,sigconf]{acmart}
\usepackage{latexsym,amsfonts,amsmath,amssymb,amsthm}
\usepackage{xspace}
\usepackage{enumitem}           

\usepackage[strict]{changepage} 
\usepackage{framed} 

\usepackage{msc}
\usepackage{anb}
\usepackage{booktabs}           
\usepackage[most]{tcolorbox}    

\usepackage{todonotes}
\usepackage{pdfcomment}

\usepackage{cleveref}

\definecolor{formalshade}{rgb}{0.95,0.95,1}
\definecolor{darkblue}{rgb}{0.0, 0.0, 0.55}
\newenvironment{formal}{%
\MakeFramed{\advance\hsize-\width\FrameRestore}%
\noindent\hspace{-4.55pt}
\begin{adjustwidth}{}{7pt}%
\vspace{2pt}\vspace{2pt}%
}
{%
\vspace{2pt}\end{adjustwidth}\endMakeFramed%
}
\newif\ifclean
\cleantrue
\newif\ifnew
\newfalse
\cleanfalse

\newcommand{\ie}{\textit{i.e.,}\xspace}
\newcommand{\eg}{\textit{e.g.,}\xspace}
\newcommand{\etc}{\textit{etc.}\xspace}
\newcommand{\exenc}{\mathsf{enc}}
\newcommand{\exdec}{\mathsf{dec}}
\newcommand{\tuple}[1]{\langle #1\rangle}

\newcommand{\aenc}{\mathsf{aenc}}
\newcommand{\mac}{\mathsf{mac}}

\newcommand{\XOR}{\oplus}
\newcommand{\sha}{\mathsf{SHA256}}
\newcommand{\KDF}{\mathsf{KDF}}
\newcommand{\KSEAF}{\mathsf{KeySeed}} 
\newcommand{\KSEAFp}{\mathsf{KeySeed'}} 
\newcommand{\CHALLENGE}{\mathsf{Challenge}} 
\newcommand{\five}{\mathsf{f5}}
\newcommand{\fives}{\mathsf{f5}^*}
\newcommand{\four}{\mathsf{f4}}
\newcommand{\three}{\mathsf{f3}}
\newcommand{\two}{\mathsf{f2}}
\newcommand{\one}{\mathsf{f1}}
\newcommand{\ones}{\mathsf{f1}^*}

\newcommand{\m}[1]{\textit{#1}} 
\newcommand{\cst}[1]{\texttt{'#1'}} 
\newcommand{\s}{$^*$}               
\newcommand{\imsi}{\m{IMSI}\xspace}
\newcommand{\supi}{\m{SUPI}\xspace}
\newcommand{\suci}{\m{SUCI}\xspace}
\newcommand{\SNname}{\m{SNname}\xspace}
\newcommand{\SNid}{\m{SNid\xspace}}

\renewcommand{\k}{K}
\newcommand{\kauth}{K_{\mathrm{aut}}}
\newcommand{\skHN}{sk_\mathrm{HN}}
\newcommand{\pkHN}{pk_\mathrm{HN}}
\newcommand{\idHN}{id_\mathrm{HN}}
\renewcommand{\r}{R}
\newcommand{\kseaf}{K_{\mathrm{SEAF}}}
\newcommand{\kausf}{K_{\mathrm{AUSF}}}

\newcommand{\sqn}{\m{SQN}\xspace}

\newcommand{\sqnUE}{\sqn_{\mathrm{UE}}}
\newcommand{\sqnHN}{\sqn_{\mathrm{HN}}}
\newcommand{\sqns}{\m{SQNs}\xspace}
\newcommand{\AUTN}{\m{AUTN}}
\newcommand{\AUTS}{\m{AUTS}}
\newcommand{\AK}{\m{AK}}
\newcommand{\AKS}{\m{AK\s}}

\newcommand{\CONC}{\m{CONC}}
\newcommand{\MAC}{\m{MAC}}
\newcommand{\MACS}{\m{MACS}}
\newcommand{\HN}{\m{HN}\xspace}
\newcommand{\HNs}{\m{HNs}\xspace}
\newcommand{\SN}{\m{SN}\xspace}
\newcommand{\SNs}{\m{SNs}\xspace}
\newcommand{\UE}{\m{UE}\xspace}
\newcommand{\UEs}{\m{UEs}\xspace}

\newcommand{\USIM}{USIM\xspace}
\newcommand{\USIMs}{USIMs\xspace}

\newcommand{\spec}[2]{\normalfont[\texttt{#1},\,\textsf{Sec.\,#2}]}  
\newcommand{\specTS}[1]{\spec{TS\,33.501}{#1}}  
\newcommand{\specTSnosec}{[\texttt{TS\,33.501}]}  
\newcommand{\SpecTS}{\texttt{3GPP\,TS\,33.501}}  
\renewcommand{\quote}[3]{\begin{formal}\textcolor{darkblue}{\bf [\texttt{#1},\,\textsf{Sec.\,#2}]} #3\end{formal}}  
\newcommand{\quoteTS}[2]{\quote{TS\,33.501}{#1}{#2}}  
\newcommand{\prop}[1]{\textcolor{violet}{\textsf{\textit{#1}}}}

\newcommand{\messLabel}[1]{\small{#1}}

\newcommand{\attack}{\includegraphics[height=5pt]{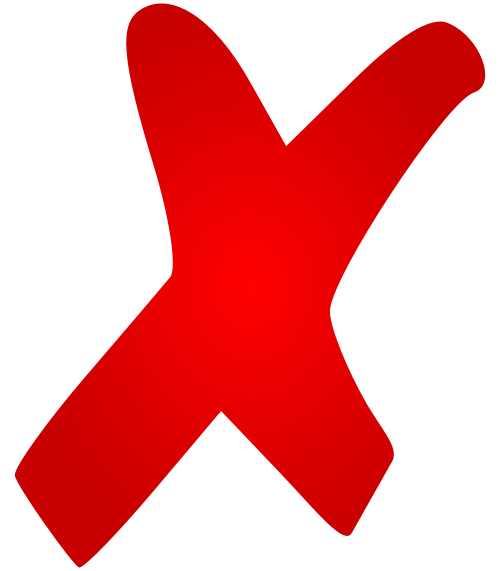}}

\newcommand{\tamarin}{\textsc{Tamarin}\xspace}
\newcommand{\proverif}{\textsc{ProVerif}\xspace}
\newcommand{\deepsec}{\textsc{DeepSec}\xspace}

\newcommand{\rwrleft}{\mathrel{-\!\!\!\!-\!\!\![}}
\newcommand{\rwrright}{\mathrel{]\!\!\!\rightarrow}}
\newcommand{\rwr}[1][]{{\mathrel{\rwrleft #1 \rwrright}}}

\newcommand{\factStyle}[1]{\textsf{#1}}

\usepackage{booktabs} 

\usepackage{balance}

\begin{document}

\fancyhf{} 
\fancyhead[R]{D. Basin, J. Dreier, L. Hirschi, S. Radomirovi{\'c}, R. Sasse and V. Stettler} 
\fancyhead[L]{A Formal Analysis of 5G Authentication}
\fancyfoot[C]{\thepage}
\setcopyright{none} 
\acmConference[]{} 
\acmYear{2018}
\settopmatter{printacmref=false, printccs=false, printfolios=false} 


\title{A Formal Analysis of 5G Authentication}
\author{David Basin}
\affiliation{
  Department of Computer Science\\
  ETH Zurich\\
  Switzerland}
\email{basin@inf.ethz.ch}
\author{Jannik Dreier}
\affiliation{Universite de Lorraine\\
  CNRS, Inria, LORIA\\
  Nancy, France}
\email{jannik.dreier@loria.fr}
\author{Lucca Hirschi}
\affiliation{
  Department of Computer Science\\
  ETH Zurich\\
  Switzerland}
\email{lucca.hirschi@inf.ethz.ch}
\author{Sa\v{s}a Radomirovi{\'c}}
\affiliation{
  School of Science and Engineering\\
  University of Dundee\\
  UK}
\email{s.radomirovic@dundee.ac.uk}
\author{Ralf Sasse}
\affiliation{
  Department of Computer Science\\
  ETH Zurich\\
  Switzerland}
\email{ralf.sasse@inf.ethz.ch}
\author{Vincent Stettler}
\affiliation{
  Department of Computer Science\\
  ETH Zurich\\
  Switzerland}
\email{svincent@student.ethz.ch}
\renewcommand{\shortauthors}{D. Basin, J. Dreier, L. Hirschi, S. Radomirovi{\'c}, R. Sasse and V. Stettler}

\begin{abstract}
Mobile communication networks connect much of the world's
population. The security of users' calls, SMSs, and mobile data
depends on the guarantees provided by the Authenticated Key Exchange
protocols used. For the next-generation network (5G), the 3GPP group
has standardized the 5G AKA protocol for this purpose.

We provide the first comprehensive formal model of a protocol from the
AKA family: 5G AKA.  We also extract precise requirements from the
3GPP standards defining 5G and we identify missing security
goals. Using the security protocol verification tool Tamarin, we
conduct a full, systematic, security evaluation of the model with
respect to the 5G security goals. Our automated analysis
identifies the minimal security assumptions required for each security
goal and we find that some critical security goals are not met, except
under additional assumptions missing from the standard. Finally, we
make explicit recommendations with provably secure fixes for the
attacks and weaknesses we found.

\end{abstract}




\maketitle

\section{Introduction}

Two thirds of the
world's population, roughly 5 billion people, are mobile
subscribers~\cite{gsma-stats}.  They
are connected to the mobile network 
via their USIM cards and are 
protected by security mechanisms
standardized by the 3rd Generation Partnership Project (3GPP) group.
Both subscribers and carriers expect security guarantees from the
mechanisms used, such as the confidentiality of user data (\eg voice
and SMS)  and that subscribers are billed only for the services
they consume.  Moreover, these properties should hold 
in an adversarial environment with malicious base stations and users.

One of the most important security mechanisms in place aims at mutually
authenticating subscribers and their carriers and establishing a
secure channel to protect subsequent communication.  For  
network generations (3G and 4G)
introduced since the year 2000, this is achieved using variants
of the Authentication and Key Agreement (AKA) protocol, standardized
by the
3GPP.  These protocols involve the subscribers, the Serving Networks
(\SNs) that have base stations in subscribers' vicinity, and Home
Networks (\HNs) that correspond to the subscribers' carriers.  The
protocols aim to enable the subscribers and the \HNs to mutually authenticate
each other and to let the subscribers and the \SNs establish a session key.
%
%

\paragraph{Next-Generation (5G)}
Since 2016, the 3GPP group has been standardizing
the next generation of mobile communication (5G)
with the aim of increasing network throughput and 
offering an  ambitious infrastructure encompassing 
new use cases.
The 5G standard will be deployed in two phases.
The first phase (Release 15, June 2018) addresses the
most critical requirements needed for commercial deployment
and forms the basis for the first deployment.
The second phase (Release 16, to be completed by the end of 2019) will
address all remaining requirements.

In June 2018, the 3GPP published the {\em final} version \texttt{v15.1.0} of Release \texttt{15} of
the Technical Specification (TS) defining the 5G security architecture and procedures~\cite{3gpp-33501}.
The authentication in 5G Release 15 is based on new 
versions of the AKA protocols, notably the new \emph{5G AKA} protocol,
which enhances the AKA protocol currently used in 4G (EPS AKA)
and which supposedly provides improved security guarantees.
This raises the following question:
\emph{What are the security guarantees
that 5G AKA actually provides and under which threat model and security assumptions?}

\paragraph{Formal Methods}
In this paper, we give a precise answer to the above question.
Namely, we apply formal methods and automated verification in the 
\emph{symbolic model} to determine precisely which
security guarantees are met by 5G AKA.
Formal methods have already proved extremely valuable in assessing the
security of large-scale, real-world security protocols such as
TLS 1.3~\cite{cremers2016automated,TLS-PV-CP,TLS-Cas-CCS17}, 
messaging protocols~\cite{Signal-PV-CP}, 
and entity authentication protocols~\cite{basin2013provably}. 
%
%
Symbolic approaches, in particular, allow one to automate reasoning
using techniques including model-checking, resolution, and rewriting.
Examples of mature verification
tools along these lines are \tamarin~\cite{TAMARIN-CAV}, \proverif~\cite{PROVERIF}, and \deepsec~\cite{DEEPSEC}.

Unfortunately, the AKA protocols, and \emph{a fortiori} 5G AKA, feature
a combination of properties that are extremely challenging for
state-of-the-art verification techniques and tools and, until very
recently, a detailed formalization was outside of their scope. First, the
flow and the state-machines of these protocols are large and complex.
%
This is due in part to the use of sequence numbers ($\sqn$) and the
need for a re-synchronization mechanism should counters become
out-of-sync.
This complexity is problematic
for tools that reason about a bounded number of sessions as they scale
poorly here. It also eliminates the option of machine-checked manual
proofs as the number of interactions is too large for humans to explore.  Second,
these protocols are stateful (the $\sqn$ counters are mutable and
persist over multiple sessions) and have numerous loops. This makes
inductive reasoning necessary and rules out fully automated tools, 
which are not yet capable of automatically finding appropriate inductive
invariants.  
Finally, the AKA protocols use the Exclusive-OR (XOR) primitive to conceal some values. This primitive
is notoriously hard to reason about  symbolically, due to its algebraic properties (\ie associativity,
commutativity, cancellation, and neutral element).
For this reason, prior works provided only limited models of the AKA
protocols, which were insufficiently precise for a satisfactory analysis;
see the discussion on related work below.
%
Given these features, we are left with just the verifier
\tamarin~\cite{TAMARIN-CAV}
as a suitable tool, and \tamarin has only recently been extended to handle
XOR~\cite{tamarin-xor}.
\smallskip{}

\noindent\textbf{Contributions.}
We describe next our three main contributions:  our formalization, models, and analysis results.

\paragraph{Formalization of the 5G Standard} 
We extract and formally interpret the standard's security assumptions
and goals.   In doing so, we identify key missing security goals and 
flaws in the stated goals.  We target a wide range of
properties --- confidentiality, authentication, and privacy --- and their fine-grained variants.  As
explained in Sections~\ref{sec:spec} and \ref{sec:spec:prop}, this required
considerable analysis and interpretation of the 3GPP Technical
Specification (722 pages across 4 documents).

\paragraph{Formal Model of 5G AKA}
We tackle the aforementioned challenges to 
provide the first faithful model of an AKA protocol that 
is detailed enough for a precise security analysis and is still
amenable to automation.  As we explain in
\Cref{sec:models},
the modeling choices for formalizing our interpretation of the standard are crucial.
To support
reasoning about our model, we develop dedicated proof techniques based
on inductive lemmas and proof strategies that guide
proof search.

\paragraph{Security Evaluation of 5G AKA}
We carry out the first formal security evaluation
of 5G authentication, providing a comprehensive analysis of the 5G AKA
protocol. This includes:
\begin{itemize}
\item \emph{a formal, systematic security evaluation}: we leverage our
  model of 5G AKA to automatically identify the minimal security
  assumptions required for each security goal to hold.
  We find that some
  critical authentication properties are violated
  prior to key confirmation, which is not clearly mandated by the standard.
  Some other properties are not met, except under assumptions on the 5G ecosystem that are
  missing from the standard.
  Additionally, we show that a privacy attack (enabling traceability) is possible for an active attacker.
  See the tables in \Cref{sec:analysis:discussion} for details.

\item \emph{recommendations}: \looseness=-1 we make explicit recommendations and
propose provably secure fixes for the attacks and weaknesses
  we identified. Most of our recommendations generalize to 5G
  Authentication as a whole, and not just 5G AKA.
\end{itemize}
\looseness=-1
We believe that our model of 5G AKA provides a valuable tool to
accompany the 5G standard's evolution and assess the security of
future proposal updates and the standard's evolution (\eg 5G phase 2).
Our model can also serve as the basis for a
comprehensive formal comparison between AKA protocols from all
generations, providing precise answers to questions like ``what
guarantees does one obtain, or lose, when moving from 4G to 5G?''
\smallskip{}

\noindent\textbf{Related Work.}
\label{sec:intro:related}
Formal methods have been applied to AKA protocols in the past, but prior work provided
 only weak guarantees due to the use of strong abstractions, protocol simplifications, 
and limitations in the analyzed properties.

The initial AKA protocol specified for 3G was manually verified by the
3GPP using
TLA and an enhanced BAN logic~\cite{specFormalAnalysis}.
The TLA analysis focused on functional properties, like the protocol recovers from
de-synchronization.
The short pen and paper proof, which was given in an enhanced BAN logic, provides
weak guarantees, \eg about key agreement and confidentiality, due to the logic's limitations. In particular, the logic does not account for, \eg compromised agents and type-flaws,
and it has had soundness issues in the past~\cite{boyd1993limitation}.
Moreover, the proof considered a simplified protocol without $\sqn$ concealment or re-synchronization
as $\sqns$ were always assumed to be synchronized.
This misses, for example, the privacy attack based on the
desynchronization error message that we observed.

\looseness=-1
\proverif has also been used to formally check untraceability and basic authentication properties of
simplified AKA protocols~\cite{PHB-sp17,arapinis2012new}.
These prior works acknowledge the challenges of formally verifying AKA protocols
but only offered limited solutions.
For instance,  the $\sqn$ counters were abstracted away by nonces
that are initially shared by \HNs and subscribers,
thus reducing the protocol to a stateless protocol.
The re-synchronization procedure was also omitted.
The \SNs and \HNs were merged into a single entity.
Furthermore, XOR was either not modeled or was replaced by
a different construct with simpler algebraic properties.
The resulting protocol was thus overly simplified and corresponding analyses
would have missed the attacks we obtain in this paper (\Cref{fig:results:auth}).
Moreover, the only authentication property that was checked is mutual aliveness between subscribers and the network.


More recently, \cite{hussain2018lteinspector} proposed a model-based testing approach that used
\proverif to carry out some analyses of EPS AKA from 4G. However, in addition to using the same aforementioned
abstractions and simplifications, they only used \proverif
to check
if specific trace executions correspond to attack traces.
%

\looseness=-1
In summary, in stark contrast to previous work, we
provide the first faithful formalization of an AKA protocol. Namely,
we formalize the entire protocol logic including the full protocol
state machine with all message flows and symbolic abstractions of all
cryptographic operators. This allows for the first comprehensive
formal analysis that characterizes the properties that are achieved in
different adversarial settings.
\smallskip{}

\noindent\textbf{Outline.}
We present in \Cref{sec:spec} the cellular network architecture and how authentication is
achieved in the 5G ecosystem using the 5G AKA protocol. We
carry out a systematic formalization of the security assumptions and
goals of the standard in \Cref{sec:spec:prop} and highlight
shortcomings. In \Cref{sec:models} we explain the basics of the
\tamarin verifier and our modeling and design choices.  We
present our comprehensive security analysis of 5G AKA and our recommendations
in \Cref{sec:analysis}.  We draw conclusions in \Cref{sec:conclusion}.

\section{5G Authentication Protocols}
\label{sec:spec}

\looseness=-1
We explain in this section how authentication and key establishment are achieved in the 5G ecosystem,
following as closely as possible the specification
\SpecTS~\cite{3gpp-33501}, referred from here on as \specTSnosec.
We simplify terminology to improve readability and refer the knowledgeable reader to the correspondence
table with the terminology from 3GPP given in \Cref{ap:index}.
We first present the general architecture and afterwards the authentication protocols.

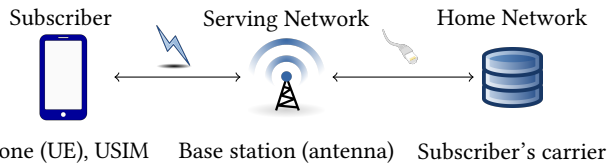
\begin{figure}[b]
  \centering
  \begin{tikzpicture}
\node (st)  at (0,-0.2) {Subscriber};
\node (s)   at (0,-1) {\includegraphics[width=1.2cm]{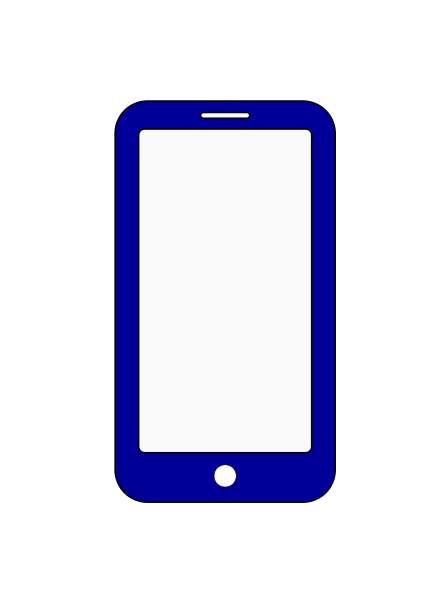}};
\node (st2) at (0,-2) {Phone (UE), USIM};

\node (snt) at (3,-0.25) {Serving Network};
\node (sn)  at (3,-1) {\includegraphics[width=1cm]{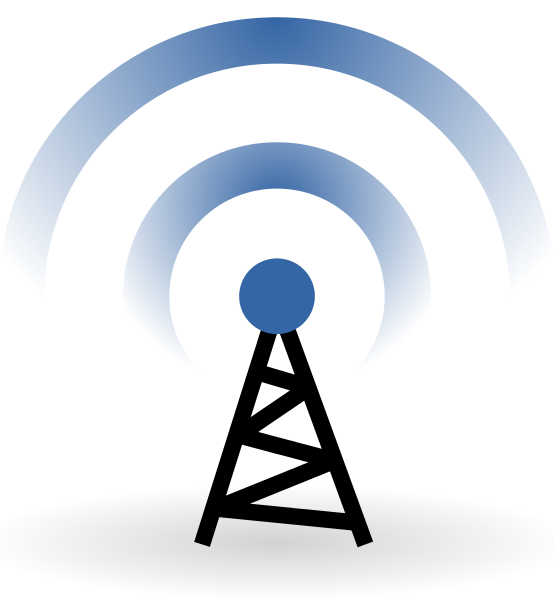}};
\node (snt2) at (3,-2) {Base station (antenna)};

\node (hnt) at (6,-0.2) {Home Network};
\node (hn)  at (6,-1) {\includegraphics[width=0.8cm]{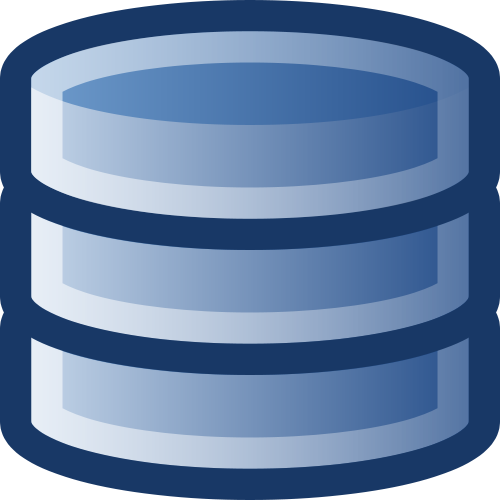}};
\node (hnt2) at (6,-1.98) {Subscriber's carrier};

\draw[<->] (s.east) to (sn.west);

\node (wl)  at (1.5,-0.6) {\includegraphics[width=0.5cm]{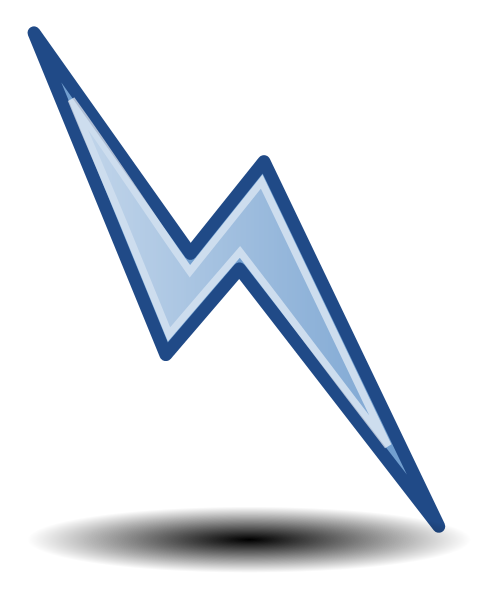}};

\draw[<->] (sn.east) to (hn.west);

\node (el)  at (4.5,-0.6) {\includegraphics[width=0.5cm]{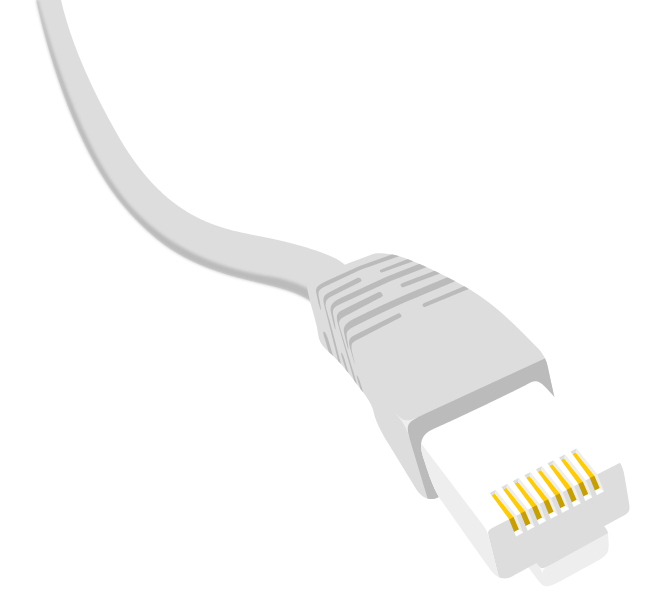}};
\end{tikzpicture}
\vspace{-15pt}
  \caption{\looseness=-1
    Overall architecture: The subscriber uses his phone (UE), equipped with a USIM, to communicate with a base station run by the \SN over an insecure wireless channel.
  The \SN communicates with the subscriber's carrier (\HN) on the right over an authenticated (wired) channel.}
  \label{fig:archi}
\end{figure}

\subsection{Architecture}
\label{sec:spec:archi}
Three main entities are involved in the cellular network architecture (see \Cref{fig:archi}).
First, \emph{User Equipment} (\UE), 
typically smartphones or IoT devices containing a \emph{Universal 
  Subscriber Identity Module} (\USIM),
are carried by \emph{subscribers}.
We shall call a \emph{subscriber} the combination of a \UE with its \USIM.
Second, \emph{Home Networks} (\HNs) contain a database of their subscribers and are responsible for their authentication.
However, subscribers may be in locations where their corresponding \HN has no \emph{base station}
(\ie antennas which may connect \UEs to the network), for example when roaming.
Therefore, the architecture has a third entity:
the \emph{Serving Networks} (\SNs) to which \UEs may attach to.
An \SN provides services (\eg call or SMS) once both the \UE and
the \SN have mutually authenticated each other (this supports billing)
and have established a secure channel with the help of the
subscriber's \HN.
The \UE and \SN communicate over the air, while the \SN and \HN communicate over an authenticated channel (we list security assumptions later in this section).

\definecolor{gris}{gray}{0.85}
\begin{figure*}[th]
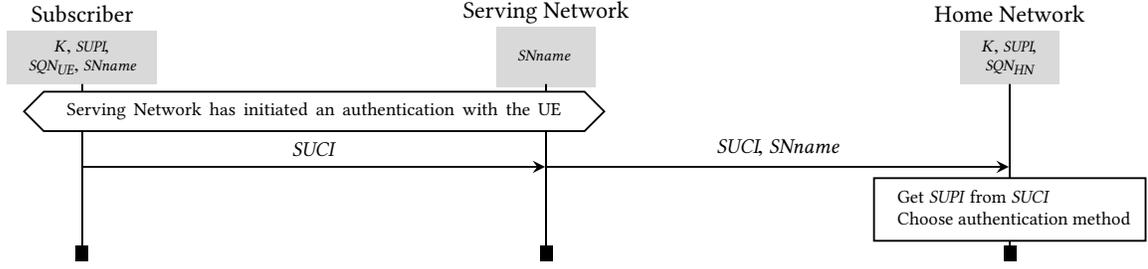

  \centering
  \setmsckeyword{}
  \drawframe{no}    
  \begin{msc}[
    /msc/title top distance=0cm,
    /msc/first level height=.1cm,
    /msc/last level height=0.4cm,
    /msc/head height=0cm,
    /msc/instance width=0cm,
    /msc/head top distance=0.5cm,
    /msc/foot distance=-0.0cm,
    /msc/instance width=0cm,
    /msc/condition height=0.2cm
    ]{}
    \setlength{\instwidth}{0\mscunit} 
    \setlength{\instdist}{6cm}  

    \declinst{UE}{              
      \begin{tabular}[c]{c}
        Subscriber \\ 
        \colorbox{gris}{\scriptsize{$\begin{array}{c}\k, \supi, \\ \sqn_{\UE}, \SNname \end{array}$}} 
      \end{tabular}
    }{}
    \declinst{SN}{              
      \begin{tabular}[c]{c}
        Serving Network \\
        \colorbox{gris}{\scriptsize{$\begin{array}{c}\raisebox{-5pt}{\SNname}\\ \phantom{k, \supi}\end{array}$}}
      \end{tabular}
    }{}
    \declinst{HN}{              
      \begin{tabular}[c]{c}
        Home Network \\ 
        \colorbox{gris}{\scriptsize{$\begin{array}{c}\k, \supi, \\ \sqn_{\HN}\end{array}$}} 
      \end{tabular}
    }{} 

    \condition{\messLabel{{\footnotesize Serving Network has initiated an authentication with the UE}}}{UE,SN}
    \nextlevel[2]
    \mess{\messLabel{\suci}}{UE}{SN}
    \mess{\messLabel{$\suci, \SNname$}}{SN}{HN}
    \nextlevel[0.3]
    \action*{\parbox{95pt}{\footnotesize{
          $\begin{array}[c]{l}
             \text{Get \supi from \suci}\\
             \text{Choose authentication method}\\
           \end{array}$
         }}}{HN}
     \nextlevel[1.0]
   \end{msc}
   \vspace{-15pt}
  \caption{Initiation of Authentication}
  \label{fig:initAuth}
\end{figure*}

As mentioned earlier, each subscriber has a \USIM with cryptographic
capabilities (\eg symmetric encryption, MAC).  Relevant for our work
is that the \USIM stores:
\begin{itemize}
\item a unique and permanent subscriber \emph{identity}, called the \emph{Subscription Permanent Identifier} (\supi),
\item the public asymmetric key $\pkHN$ of its corresponding \HN,
\item a long-term \emph{symmetric key}, denoted as $\k$ (used as a shared secret between subscribers and their corresponding \HNs), and
\item a counter, called \emph{Sequence Number}, denoted as $\sqn$.
\end{itemize}
The \HN, associated to some subscriber, stores the same information in its database.

\paragraph{Simplifications.}
In the standard, \SNs and \HNs are composed of several sub-entities (\eg \HNs consist of a database, authentication server, \etc).
However, very few security properties require this level of granularity.
We thus have chosen to consider these three larger logical entities (see \Cref{ap:index} for more details).


\subsection{Authentication Protocols}
\label{sec:spec:protos}
To enable \SNs and subscribers to establish secure channels and
authenticate each other, the 3GPP has specified two
authentication methods: \emph{5G AKA} and \emph{EAP-AKA'}. The choice between those two methods is left to the \HN,
once it has correctly identified the subscriber with the \emph{Initialization Protocol}. We now describe these three security
protocols. (All cryptographic messages are precisely described in \Cref{ap:index}.)


\subsubsection{Initialization Protocol \specTS{6.1.2}}
\Cref{fig:initAuth} depicts the sub-protocol responsible for the
subscribers' identification and initializing the
authentication.  Once the \SN has triggered an authentication with the
subscriber, the latter sends a randomized encryption of the $\supi$ (for
privacy reasons, as we explain in \Cref{sec:spec:prop:privacy}):
$\suci= \tuple{\aenc(\tuple{\supi, \r_s}, \pkHN), \m{idHN}}$, where
$\aenc(\cdot)$ denotes asymmetric encryption, $\r_s$ is a random
nonce, and \m{idHN} uniquely identifies an \HN.
The identifier \m{idHN} enables the \SN to request authentication material from
the appropriate \HN.
Upon reception of the $\suci$ along with the \SN's identity (referred to as \SNname),
the \HN can retrieve the \supi, the subscribers' identity, and choose an authentication method.
Note that $\supi$ also contains $\m{idHN}$ and therefore identifies both a subscriber and its \HN.
%
%
\smallskip{}

\subsubsection{The 5G AKA Protocol \specTS{6.1.3.2}}
\label{sec:proto:5G-AKA}
As mentioned before, the key $K$ is used as a long-term shared secret, and $\sqn$ provides
replay protection\footnote{This design choice is for historical
  reasons: old \USIMs (\eg in 3G and  4G)
  did not have the capability to generate random nonces.} for the subscriber. While $\sqn$ should
be synchronized between the subscriber and the \HN, it may happen that
they become out-of-sync, \eg due to message loss.
We thus use $\sqnUE$ (respectively $\sqnHN$) to refer to the SQN value stored in the \UE (respectively \HN).
The 5G-AKA protocol consists of two main phases: a \emph{
  challenge-response} and an optional
\emph{re-synchronization procedure} (that updates the SQN on the \HN side
in case the \sqn is out of-sync).
The entire 5G AKA protocol flow is depicted in \Cref{fig:5GAKA}.
\smallskip{}

\noindent{{\it Challenge-Response.}}
Upon receiving a request for authentication material,
the \HN computes an \emph{authentication challenge} built from:
\begin{itemize}
\item a random nonce $R$ (the challenge),
\item $\AUTN$ (proving the challenge's freshness and authenticity),
\item \m{HXRES\s} (response to the challenge that \SN expects),
\item $\kseaf$ (key seed for the secure channel that the subscriber and \SN will eventually establish).
\end{itemize}
The functions $\mathsf{f1}-\mathsf{f5}$, used to compute the 
authentication parameters, are one-way keyed cryptographic functions completely 
unrelated with each other, and $\oplus$ denotes Exclusive-OR.
$\CHALLENGE(\cdot)$ and $\KSEAF(\cdot)$ are complex Key Derivation Functions (KDFs);
see \Cref{ap:index} for more details.
$\AUTN$ contains a Message Authentication Code (MAC) of the concatenation of 
$R$ with the corresponding sequence number $\sqnHN$ stored for this subscriber.
A new sequence number is generated by incrementing the counter.
The sequence number $\sqnHN$ allows the subscriber to verify 
the freshness of the authentication request to defend against replay attacks
and the MAC proves the challenge's authenticity.
The \HN does not send the challenge's full response \m{RES\s} to the \SN but only
a hash therereof; the rationale being that \HNs are willing to have assurance of
the presence of its subscribers even with malicious \SNs.

The \SN stores $\kseaf$ and the challenge's expected response and then forwards
the challenge to the subscriber.  Upon receiving the challenge,
the subscriber first checks its authenticity and freshness.
To do this, the subscriber extracts $x\sqnHN$ and $MAC$
from $\AUTN$ and checks that:
\begin{enumerate}[label=(\roman*)]
\item $MAC$ is a correct MAC value with respect to~$\k$, and replies\linebreak[4]
  $\cst{Mac\_{failure}}$ if it is not the case,
\item the authentication request is fresh\footnote{The freshness check may also
consider non-normative protection against the wrapping around of $\sqnHN$
which we do not describe here; see \spec{TS\,33.102}{C}.},
  \ie $\sqnUE < x\sqnHN$,
  and replies 
  $\tuple{\cst{Sync\_{failure}},\AUTS}$ otherwise ($\AUTS$ is
  explained in the 
re-synchronization procedure below).
\end{enumerate}
If all checks hold, then the subscriber computes the
key seed $\kseaf$, which is used to secure subsequent messages.
It also computes the 
authentication response $\m{RES\s}$ and sends it to the \SN.
The \SN checks that this response is as expected and forwards it to the \HN,
who validates it.
If this validation succeeds then the \HN confirms to the \SN that the authentication is successful and sends the \supi to the \SN.
Subsequent communications between the \SN and the subscriber can be secured using the key seed $\kseaf$.
\smallskip{}

\noindent{\it Re-synchronization~procedure \specTS{6.1.3.2.1}.}
In case of a synchronization failure (case \texttt{$\neg$(ii)}),
the subscriber replies with $\tuple{\cst{Sync\_{failure}},\AUTS}$. The
$\AUTS$ message enables
 the \HN to re-synchronize with the subscriber by replacing its own $\sqnHN$ 
by the sequence number of the subscriber $\sqnUE$;\linebreak[4] see \spec{TS\,33.102}{6.3.5,6.3.3}.
However, $\sqnUE$ is not transmitted in clear text to avoid being eavesdropped on (it is privacy sensitive
as explained in \Cref{sec:spec:prop:privacy}).
Therefore, the specification requires $\sqn$ to be \emph{concealed};
namely, it is XORed with
a value that should remain private: $\AKS=\fives(\k,\r)$.
Formally, the concealed value is $\CONC^*=\sqnUE \xor \AKS$, which
allows the \HN to extract $\sqnUE$ by computing $\AKS$. Note that $\fives$ and
$\ones$ are independent one-way keyed cryptographic functions, completely unrelated to the functions 
$\one-\five$. Finally, $\AUTS=\tuple{\CONC^*,\mathit{MAC}^*}$,
where $\mathit{MAC}^*=\ones(\k,\tuple{\sqnUE, \r})$, allowing the \HN to authenticate this message as coming from
the intended subscriber.

\subsubsection{The EAP-AKA' Protocol \specTS{6.1.3.1} and \textsf{[RFC 5448]}}
EAP-AKA' is very similar to 5G AKA: it relies on the same mechanisms
(challenge-response with $\k$ as a shared secret and $\sqn$ for replay
protection) and uses similar cryptographic messages. The
main difference is the flow and some key derivation functions 
are slightly changed.  Since we focus our analysis on the 5G AKA
authentication method, we do not describe those differences in detail
here and refer the curious reader to \Cref{ap:EAP-AKAp}.


\definecolor{gris}{gray}{0.85}
\begin{figure*}[th]
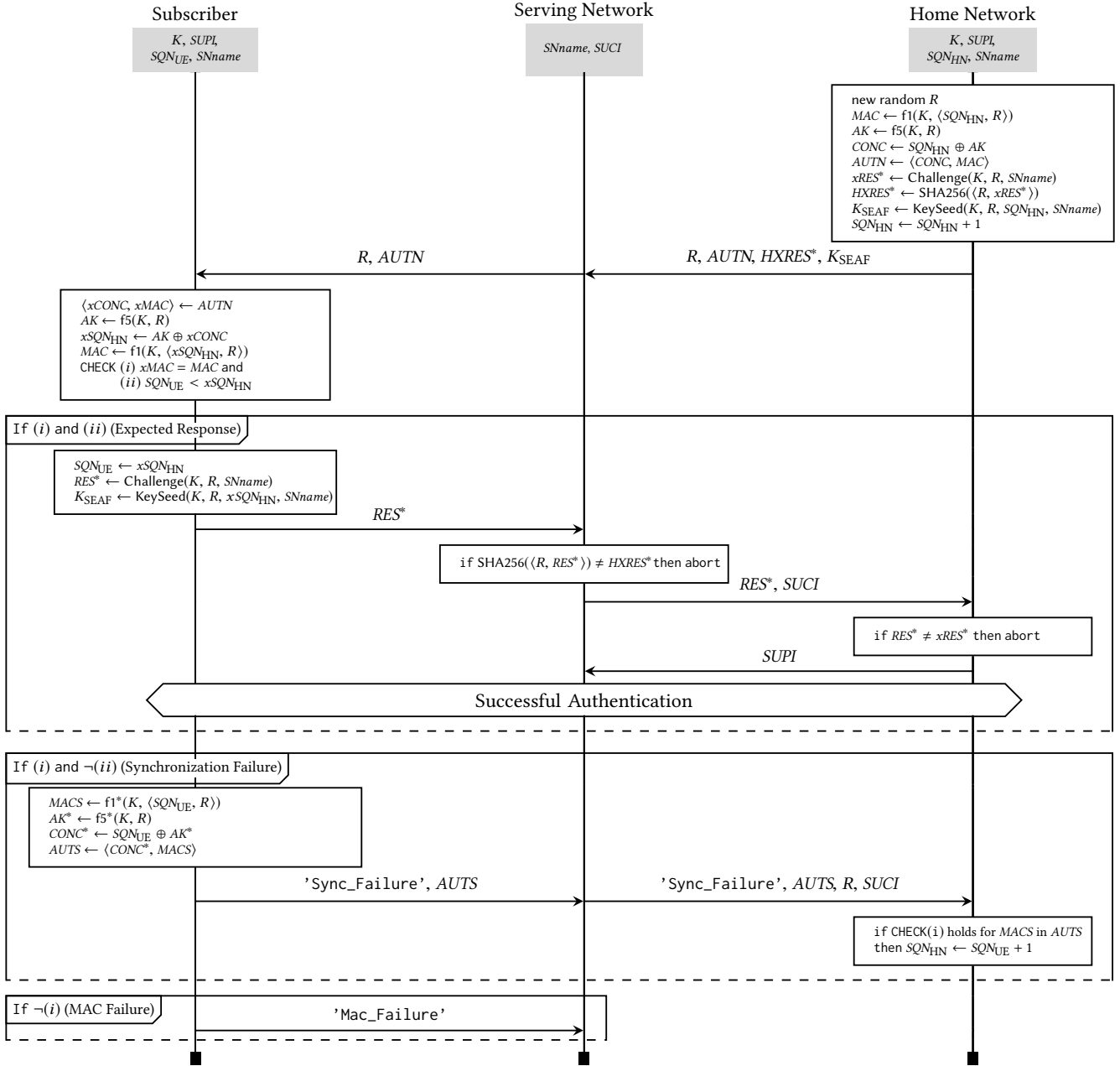

  \centering
  \setmsckeyword{}
  \drawframe{no}    
  \begin{msc}[
    /msc/title top distance=0cm,
    /msc/first level height=.2cm,
    /msc/last level height=0.2cm,
    /msc/head height=0cm,
    /msc/instance width=0cm,
    /msc/head top distance=0.5cm,
    /msc/foot distance=-0.0cm,
    /msc/instance width=0cm,
    /msc/condition height=0.2cm
    ]{}
    \setlength{\instwidth}{0\mscunit} 
    \setlength{\instdist}{6cm}  

   \declinst{UE}{              
      \begin{tabular}[c]{c}
        Subscriber \\ 
        \colorbox{gris}{\scriptsize{$\begin{array}{c}\k, \supi, \\ \sqn_{\UE}, \SNname \end{array}$}} 
      \end{tabular}
    }{}
    \declinst{SN}{              
      \begin{tabular}[c]{c}
        Serving Network \\
        \colorbox{gris}{\scriptsize{$\begin{array}{c}\raisebox{-5pt}{\SNname, \suci}\\ \phantom{k, \supi}\end{array}$}}
      \end{tabular}
    }{}
    \declinst{HN}{              
      \begin{tabular}[c]{c}
        Home Network \\ 
        \colorbox{gris}{\scriptsize{$\begin{array}{c}\k, \supi, \\ \sqn_{\HN}, \SNname \end{array}$}} 
      \end{tabular}
    }{} 

    \action*{\parbox{4.2cm}{\scriptsize{$
          \begin{array}[c]{l}
            \mathsf{new~random}\ \r \\
            \MAC\leftarrow \one(\k,\tuple{\sqnHN, \r}) \\
            \AK\leftarrow \five(\k,\r) \\
            \CONC\leftarrow \sqnHN \xor \AK \\
            \AUTN\leftarrow  \tuple{\CONC, \MAC} \\
            \m{xRES\s}\leftarrow \CHALLENGE(\k,\r,\SNname) \\
            \m{HXRES\s}\leftarrow \sha(\tuple{\r,\m{xRES\s}}) \\            
            \kseaf \leftarrow \KSEAF(\k,\r,\sqnHN,\SNname) \\
            \sqnHN \leftarrow \sqnHN + 1
          \end{array}
          $}}}{HN}
    \nextlevel[6]
    \mess{\messLabel{$\r,\AUTN, \m{HXRES\s}, \kseaf$}}{HN}{SN}
    \mess{\small{$\r,\AUTN$}}{SN}{UE}
    \nextlevel[0.5]

    \action*{\parbox{4.0cm}{\scriptsize{$
          \begin{array}[c]{l}
            \tuple{\m{xCONC}, \m{xMAC}} \leftarrow \AUTN\\
            \AK\leftarrow \five(\k,\r) \\
            \m{x}\sqnHN \leftarrow \AK\xor \m{xCONC}\\
            \MAC\leftarrow  \one(\k, \tuple{\m{x}\sqnHN, \r}) \\
            \mathtt{CHECK}\; (i)\; \m{xMAC}=\MAC \;\mathtt{and}\;\\
            \phantom{\mathtt{CHECK}}\;  (ii)\; \sqnUE < \m{x}\sqnHN  
            \\
          \end{array}
          $}}}{UE}
    \nextlevel[4]

    \inlinestart[3.0cm][63pt]{thenThen}{\footnotesize $\mathtt{If}\;(i)\;\mathtt{and}\;(ii)$
      {\setlength{\fboxsep}{0pt}\colorbox{white!30}{(Expected Response)}}}{UE}{HN}
    \nextlevel[1.1]
    \action*{\parbox{4.2cm}{\scriptsize{$
          \begin{array}[c]{l}
            \sqnUE \leftarrow \m{x}\sqnHN\\
            \m{RES\s}\leftarrow \CHALLENGE(\k,\r,\SNname) \\
            \kseaf \leftarrow \KSEAF(\k,\r,x\sqnHN,\SNname) \\
          \end{array}
          $}}}{UE}
    \nextlevel[2.5]
    \mess{\small{$\m{RES\s}$}}{UE}{SN}
    \nextlevel[0.5]
    \action*{\parbox{4.3cm}{\scriptsize{$
          \begin{array}[c]{l}
            \mathtt{if}\  \sha(\tuple{\r,\m{RES\s}}) \neq \m{HXRES\s}
            \mathtt{then}\ \mathtt{abort}
          \end{array}
          $}}}{SN}
    \nextlevel[1.8]
    \mess{\small{$\m{RES\s}, \suci$}}{SN}{HN}
    \nextlevel[0.5]
    \action*{\parbox{3.5cm}{\scriptsize{$
          \begin{array}[c]{l}
            \mathtt{if}\  \m{RES\s} \neq \m{xRES\s}\ 
            \mathtt{then}\ \mathtt{abort}
          \end{array}
          $}}}{HN}   
    \nextlevel[1.7]
    \mess{\small{$\supi$}}{HN}{SN}
    \nextlevel[0.4]
    \condition{Successful Authentication}{UE,SN,HN}
    \nextlevel[1.5]
    \inlineend*{thenThen}
    \nextlevel[0.7]

    \inlinestart[3.0cm][63pt]{thenElse}{\footnotesize $\mathtt{If}\;(i)\;\mathtt{and}\;\lnot(ii)${\setlength{\fboxsep}{0pt} \colorbox{white!30}{\strut(Synchronization Failure)}}}{UE}{HN}
    \nextlevel[1.1]
    \action*{\parbox{5.0cm}{\scriptsize{$
          \begin{array}[c]{l}
            \MACS \leftarrow \ones(\k,\tuple{\sqnUE, \r}) \\
            \AKS \leftarrow \fives(\k,\r) \\
            \m{CONC\s} \leftarrow \sqnUE \XOR \AKS \\
            \AUTS\leftarrow \tuple{\m{CONC\s}, \MACS}\\
          \end{array}
          $}}}{UE}
    \nextlevel[3.6]
    \mess{\small{$\cst{Sync\_Failure}, \AUTS$}}{UE}{SN}
    \mess{\small{$\cst{Sync\_Failure}, \AUTS, \r, \suci$}}{SN}{HN}
    \nextlevel[0.5]
    \action*{\parbox{3.5cm}{\scriptsize{$
          \begin{array}[c]{l}
            \mathtt{if}\ \mathtt{CHECK (i)} \text{ holds for \m{MACS} in \AUTS}\\
            \mathtt{then}\ \sqnHN \leftarrow \sqnUE + 1
          \end{array}
          $}}}{HN}   
    \nextlevel[2]
    \inlineend*{thenElse}
    \nextlevel[0.5]

    \inlinestart[3.0cm][10pt]{else}{\footnotesize $\mathtt{If}\;\lnot(i)$ (MAC Failure)}{UE}{SN}
    \nextlevel[1.1]
    \mess{\small{$\cst{Mac\_Failure}$}}{UE}{SN}
    \nextlevel[0.3]
    \inlineend*{else}
  \end{msc}
  \vspace{-10pt}
  \caption{The 5G AKA protocol (continuing \Cref{fig:initAuth})}
  \label{fig:5GAKA}
\end{figure*}

\newcommand{\size}{\small}

\section{Threat Model and Security Goals}
\label{sec:spec:prop}
In this section, we derive precise, formal security goals from the informal
descriptions given in the Technical
Specification (TS) and Technical Requirement (TR) documents issued by
the 3GPP. Our formal definitions are our interpretation of these texts. 
We support them 
with quotes from and references to relevant excerpts
of the TS and TR documents. 
%
The full list of relevant excerpts
along with explanations of our interpretations is given in \Cref{ap:prop}.

The extraction of precise properties from the standard's informally stated
goals is an important prerequisite to applying a security
protocol analysis tool (like \tamarin). It is thus
a crucial step in the security analysis of a complex protocol such as
5G AKA.

\subsection{Security Assumptions and Threat Model}
\label{sec:spec:assumptions}
\subsubsection{Assumptions on Channels}
The channel between the \SN and the \HN provides
confidentiality,
integrity,
authenticity, and
replay protection \specTS{5.9.3}.

The channel between the subscribers and \SNs 
is subject to eavesdropping by passive attackers
and manipulation, interception, and injection of messages by active attackers.
A \emph{passive attacker} listens to signaling messages (\ie messages sent on the physical layer)
and can thus eavesdrop on all messages exchanged in its vicinity, but it never emits a signal.
An \emph{active attacker} sets up a fake base station to send and receive signaling messages, \eg to impersonate \SNs.
While no 5G-specific hardware is publicly available yet, we recall
that 
4G base stations have been built
using open-source and freely available software and hardware~\cite{ravi-NDSS16,Golde13}.
From now on, we shall consider active attackers, except when explicitly
stated otherwise.

\subsubsection{Assumptions on Cryptographic Primitives.}
\label{sec:spec:assumptions:primitives}
The functions $\one$, $\ones$, and $\two$ are message authentication functions,
and $\three,\four,\five,\fives$ are key derivation functions \spec{TS\,33.102}{3.2,6.3.2}.
To 
our knowledge there is no comprehensive set of standardized security requirements
for these functions. The requirements 
in \spec{TS\,33.105}{5} are
insufficient, but 
%
we infer from the informal presentation in \spec{TS\,33.102}{3.2} and requirements in \spec{TS\,33.105}{5} that the former provide only integrity protection 
and 
the latter 
both integrity and confidentiality protection. 
However, since $\one$ and $\ones$ are applied to data that should be
secret,
such as $\sqn$ (see \Cref{sec:spec:prop:privacy}),
it is our understanding that they should also 
preserve the confidentiality of their inputs.
%
We therefore assume in our analysis that all these functions protect
both integrity and confidentiality, but we stress that this is either
underspecified or subscribers' privacy is put at risk (see \Cref{sec:spec:prop:privacy}).



\subsubsection{Assumptions on Parties}
To provide strong, fine-grained guarantees, we consider different compromise scenarios.
First, we consider an attacker who can compromise some \SNs.
This means that the attacker gets access to an authenticated channel between the compromised \SN and
\HNs, which he can use to eavesdrop on and inject messages.
This is a reasonable assumption in 5G, where authentication methods should provide security guarantees even in the presence
of genuine but malicious \SNs \specTS{6.1.4.1}. In such situations, the \HNs may cooperate with such \SNs to authenticate some subscriber.
In practice, this may happen in roaming situations.
Next, we consider that the attacker may have genuine \USIMs and compromised \USIMs under its control.
For those compromised subscribers, the attacker can access all secret values stored in the \USIMs; \ie $\supi$, $\k$, and
$\sqn$.
Finally, the attacker can access all long-term secrets, $\k,\skHN$,
and $\supi$, from compromised \HNs.

\subsubsection{Assumptions on Data Protection}
The subscriber credentials, notably the key $\k$ and the identifier $\supi$, shared between subscribers and \HNs, should initially be secret, provided they belong to non-compromised agents \specTS{3.1}.

The sequence number $\sqn$ is a 48-bit counter or a 43-bit counter \spec{TS\,33.102}{6.3.7,C.3.2}
and therefore guessable with a very low probability.
Note that an offline guessing attack on the sequence number counter is
not possible, and online attacks on the \UE first require a correct MAC
(based on the shared secret $\k$) before the \UE responds whether the
$\sqn$ was acceptable.
We thus consider a reasonable threat model where the value of $\sqn$ is unknown
to the attacker when the attack starts, but the attacker knows how it is
incremented during the attack.
This corresponds to an attacker who
(i) can monitor the activity of targeted subscribers in its vicinity during the attack but
(ii) can neither guess the initial value of $\sqn$
(iii) nor can he monitor targeted subscribers all the time (\ie from their first use of the \USIM up to the attack time).

While not explicitly stated in the specification, we shall assume that the private
asymmetric key $\skHN$ is initially secret.

\subsection{Security Requirements}
\label{sec:prop:goals}
We now extract and interpret from the 5G documents the security goals that  5G AKA should achieve according to the 5G standard.

\subsubsection{Authentication Properties}
\label{sec:spec:prop:auth}
The 5G specifications make
claims about authentication properties at different places in the documents. We have identified
relevant claims and translated them into formal security goals, indicated in \prop{purple, cursive text}.
We use Lowe's taxonomy of authentication properties~\cite{lowe-taxonomy} to make the goals precise, prior to formalization.
These 
properties 
are well established and understood, avoiding ambiguity~\cite{basin2015improving}.
Moreover, there is a formal relationship between the taxonomy and mathematical definitions of security properties
that can be directly modeled in \tamarin~\cite{tamarin-manual}.

We give an overview of Lowe's taxonomy and its relationship with
formal definitions of authenticity in \Cref{ap:taxonomy}.
Intuitively, the taxonomy specifies, from an agent A's point of view, four levels of authentication between two agents A and B: (i) aliveness, which only ensures that B has been running the protocol previously, but not necessarily with A; (ii) weak agreement, which ensures that B has previously been running the protocol with A, but not necessarily with the same data; (iii) non-injective agreement, which ensures that B has been running the protocol with A and both agree on the data; and (iv) injective agreement, which additionally ensures that for each run of the protocol of an agent there is a unique matching run of the other agent, and prevents replay attacks.

Note that the 5G specification considers some authentication properties to be \emph{implicit}.
This means that the guarantee is provided only after an additional \emph{key confirmation roundtrip} (with respect to $\kseaf$)
between the subscribers and the \SN. We discuss the resulting problems and critique this design choice in \Cref{sec:analysis:discussion:implicit}.

\paragraph{Authentication between subscribers and \HNs.}
First, the subscribers must have the assurance that authentication can only be successful with \SNs authorized by their \HNs;
see \specTS{ 6.1.1.3} and:
\quoteTS{5.1.2}{\size
  {\bf Serving network authorization by the home network:} Assurance
  [that the subscriber] is connected to a
  serving network that is authorized by the home network. [...]
  This authorization is `implicit' in the sense that it is implied by a successful authentication and key agreement run.
}
\noindent Formally, a \prop{subscriber must obtain non-injective agreement on $\SNname$ with its \HN after key confirmation.}

In 5G, the trust assumptions are different than in previous standards, like 3G or 4G. Most notably, the level of trust
the system needs to put into the \SNs has been reduced. One important property provided by 5G
is that an \SN can no longer fake authentication requests
with the \HNs for subscribers not attached to one of its base stations \specTS{6.1.4.1}.
Formally, \prop{the \HNs obtain the aliveness of its subscribers at that \SN,
  which is non-injective agreement  on \SNname from the \HNs' point of view with the subscribers}.

\paragraph{Authentication between subscribers and \SNs.}
As expected, the \SNs shall be able to authenticate the subscribers:
\quoteTS{5.1.2}{\size
  {\bf Subscription authentication}: The serving network shall authenticate the Subscription Permanent Identifier (SUPI) in
  the process of authentication and key agreement between UE and network.
}
\noindent Formally, \prop{the \SNs must obtain non-injective agreement on $\supi$ with the subscribers.} As $\supi$ is the subscriber's identifier this is actually
just \prop{weak agreement for the \SNs with the subscribers}.
Moreover, since $\supi$ also contains $\idHN$, an agreement on $\supi$ entails an agreement on $\idHN$.

Conversely, the subscribers shall be able to authenticate the \SNs:
\quoteTS{5.1.2}{\size
  {\bf Serving network authentication}: The UE shall authenticate the serving network identifier through implicit key authentication.
}
\noindent Since $\SNname$ is the \SN's identifier,
\prop{the subscribers must obtain weak agreement with the \SNs after key confirmation.}

\paragraph{Authentication between \SNs and \HNs.}
\looseness=-1
The \SNs shall be able to authenticate subscribers that are authorized by their corresponding \HN:
\quoteTS{5.1.2}{\size
  {\bf UE authorization}: The serving network shall authorize the UE through the subscription profile obtained from the home network.
UE authorization is based on the authenticated SUPI.
}
\noindent \prop{The \SNs must obtain non-injective agreement on $\supi$ with the \HNs}.

\subsubsection{Confidentiality Properties}
\label{sec:spec:prop:conf}
While it is not clearly specified,  obviously 5G-AKA should ensure the
\prop{secrecy of $\kseaf$, $\k$, and $\skHN$} (see similar goals in 3G~\spec{TS\,133.102}{5.1.3}).

5G-AKA should also ensure that knowledge of the session key $\kseaf$ established in one session is insufficient to deduce another session key $\kseaf'$  
established in either a previous session or  a later session \specTS{3}.
Formally,  the key \prop{$\kseaf$ established in a given session remains confidential even when the attacker
  learns the $\kseaf$ keys established in all other sessions}.
Note that this is different from \prop{forward secrecy} and \prop{post-compromise secrecy~\cite{cohn2016post}},
which fail to hold as we shall see in \Cref{sec:analysis:res}.
Forward and post-compromise secrecy require session key secrecy even 
when
long-term key material is compromised. 
5G-AKA does not meet these requirements as
knowledge of the key $\k$ allows an attacker to derive all past and
future keys.

Finally, the same key $\kseaf$ should never be established twice \spec{TS\,133.102}{6.2.3}.
This will be analyzed as part of \prop{\emph{Injective} agreement properties on the established key $\kseaf$}
for different pairs of parties.

\subsubsection{Privacy Properties}
\label{sec:spec:prop:privacy}
We first emphasize the importance given to privacy in 5G documentation:
\quote{TR\,33.899}{4.1,4.2}{\size
\looseness=-1
  Subscription privacy deals with various aspects related to the protection of subscribers' personal
  information, \eg identifiers, location, data, etc.
  [...]
  The security mechanisms defined in NextGen shall be able to be configured to protect subscriber's privacy.
}
\quote{TR\,33.899}{5.7.1}{\size
  The subscription privacy is very important area for Next Generation system as can be seen by the growing attention
  towards it, both inside and outside the 3GPP world.
  [...]
}
This important role given to privacy can be explained by numerous, critical attacks that have breached privacy (\eg with IMSI-catchers~\cite{ravi-NDSS16,Broek2015})
in previous generations; see the survey~\cite{mobileSoK17}.
We also recall that privacy was already a concern in 3G:
\quote{TS\,133.102}{5.1.1}{(3G)\size 
  ~The following security features related to user identity confidentiality are provided:
  \begin{itemize}
  \item  {\bf user identity confidentiality}: the property that the permanent user identity (IMSI) of a user to whom a services
    is delivered cannot be eavesdropped on the radio access link;
  \item  {\bf user location confidentiality}: the property that the presence or the arrival of a user in a certain area cannot be
    determined by eavesdropping on the radio access link;
  \item  {\bf user untraceability}: the property that an intruder cannot deduce whether different services are delivered to the
    same user by eavesdropping on the radio access link.
  \end{itemize}
}
Thus, 3G already had security requirements for user identity confidentiality, anonymity, and untraceability.
However, these properties are required by the standard only against a \emph{passive} attacker, i.e., one who only eavesdrops on the radio link.
We criticize this restriction in \Cref{sec:analysis:discussion:privacy}.
We now list more precise requirements on privacy in 5G.
\smallskip{}

\looseness=-1
In 5G, the $\supi$ is considered sensitive and must remain secret since it uniquely identifies users
\specTS{5.2.5,6.12}.
Indeed, an attacker who obtains this value can 
identify a subscriber, leading to classical user location attacks (see \spec{TS\,133.102}{5.1.1} above),
much like passive IMSI-catcher attacks.
Formally, \prop{the $\supi$ shall remain secret in the presence of a passive attacker}.

Similarly, the $\sqn$ must remain secret \spec{TS\,33.102}{6.2.3,\linebreak[4]C.3.2}.
An additional reason that is not explicitly stated
is that the $\sqn$ leaks the number of successful authentications the corresponding \USIM has performed
since it was manufactured, which is strongly correlated to its age and activity. This is even more
critical when the attacker learns the $\sqn$ at different times, as this allows activity estimation for that time-period.
Formally, \prop{the $\sqn$ shall remain secret in the presence of a passive attacker}.

Preventing the attacker from learning identifying data (\ie $\supi$, $\sqn$) is insufficient
 protection against privacy attacks such as traceability attacks (we show an example
in \Cref{sec:analysis:discussion:privacy}).
While no formal and explicit statement is made on the necessity of ensuring untraceability 
for 5G, several claims in TR and TS documents (see \Cref{ap:prop:privacy})
and the fact that it was required for 3G (\spec{TS\,133.102}{5.1.1}, see above),
suggest that this property is relevant for 5G as well. 
Therefore, formally, 5G authentication methods should provide \prop{untraceability of
  the subscribers in the presence of a passive attacker}.

\subsection{Security Goals are Underspecified}
\label{sec:prop:under}


We now discuss the aforementioned standardized security goals and critique the lack of precision in the standard.
We show that the requirements specified in the standard are not sufficient
to provide the expected security guarantees in the context of mobile communication telephony use cases.
This is completely independent of whether or not the proposed protocols actually fulfill these properties (which we examine in \Cref{sec:analysis}).
%

First, given that the protocol is an
\emph{Authenticated Key Exchange} protocol, we expect at least mutual authentication requirements and
agreement properties on the established key.
It is thus surprising that the standard does not require any agreement on $\kseaf$.
The different pairs of roles, especially subscribers and \SNs should at least obtain non-injective agreement on the shared key $\kseaf$.
Moreover, $\kseaf$ should be different for each session. This is a critical requirement, especially for typical use cases for these protocols.
Indeed, if this property is not provided, an attacker could make \UEs and \SNs establish a secure channel based on a key that has been previously used,
and could therefore replay user data.
The crucial missing requirements are \prop{injective agreements on
  $\kseaf$ between pairs of parties, in particular between the \SNs and subscribers.}

The standard specifies authentication properties as weak \emph{authorization} properties
that can be formalized as non-injective agreement on the roles' identifiers, or simply weak agreement properties
(see \Cref{sec:spec:prop:auth}).
We discuss the standard's restriction to ``implicit authentication'' in \Cref{sec:analysis:discussion:implicit}.
As explained earlier, 5G requires \HNs to have the assurance that \UEs are attached to \SNs
\specTS{6.1.4.1} currently.
However, a non-injective agreement on $\SNname$ from an \HN towards a subscriber is too weak
since it suffices that the subscriber has attached to the corresponding \SN in some session in the past to fulfill the property.
It is crucial for the \HNs to obtain assurance that
the subscriber is attached to the \SN during the \emph{present} session.  The
derivation of $\kseaf$ includes $\SNname$ for the binding to \SN. This
derivation also includes a nonce $\r$, from which we obtain the
desired assurance as a corollary of \prop{injective agreement on
  $\kseaf$ from the \HNs towards the subscribers}, which we consider
instead.
%

Similarly, the subscribers should have the assurance that the \SNs with which they establish secure channels
are known and trusted by their \HNs at the time of the authentication, not only in some past session.
Therefore, they 
should obtain \prop{injective agreement on $\kseaf$ (which is bound to $\SNname$) with the \HNs.}
While less critical, other pairs of roles should also have stronger assurance.
%
We describe how the standard can be improved in this regard in \Cref{sec:analysis:rec}.

\section{Formal Models}
\label{sec:models}
In this section, we give a basic introduction to the symbolic model of cryptographic protocols and the tool \tamarin that automates reasoning
in this model (\Cref{sec:formal:tamarin}). Afterwards, we give an overview on how security properties can be modeled using \tamarin (\Cref{sec:formal:prop}).
Next, after describing our modeling choices (\Cref{sec:formal:choices}), we describe the challenges associated
with modeling a large, complex protocol like 5G AKA and how we overcame them (\Cref{sec:formal:models}).

\subsection{The Tamarin Prover}
\label{sec:formal:tamarin}

To analyze 5G AKA, we used the \tamarin prover~\cite{schmidt2012automated}.
\tamarin is a state-of-the-art protocol verification tool for the \emph{symbolic model}, which supports stateful protocols, a high level of automation, and equivalence properties~\cite{tamarin-equiv}, which are necessary to model privacy properties such as unlinkability.
It has previously been applied to real-world protocols with complex state machines, numerous messages, and complex security properties such as TLS 1.3~\cite{tamarin-tls}.
Moreover, it was recently extended with support for XOR~\cite{tamarin-xor}, a key ingredient for faithfully analyzing 5G AKA.
We chose \tamarin as it is currently the only tool that combines all these features, which are essential for a detailed analysis of 5G AKA.

In the symbolic model and a fortiori in \tamarin,
messages are described as terms. For example, $\exenc(m,k)$ represents the message $m$ encrypted using the key $k$.
The algebraic properties of the cryptographic functions are then specified using equations over terms.
For example the equation $\exdec(\exenc(m,k),k) = m$ specifies the expected semantics for symmetric encryption:
the decryption using the encryption key yields the plaintext.
As is common in the symbolic model, cryptographic messages do not satisfy other properties than those intended algebraic properties,
yielding the so-called \emph{black box cryptography assumption}
(\eg one cannot exploit potential weaknesses in cryptographic primitives).

The protocol itself is described using multi-set rewrite rules.
These rules manipulate multisets of \emph{facts}, which model the current state of the system with \emph{terms} as arguments.
\begin{example}\label{ex:hashmsr}
The following rules describe a simple protocol that sends an encrypted message.
The first rule creates a new long-term shared key $k$ (the fact $\factStyle{!Ltk}$ is persistent: it can be used as a premise
multiple times).
The second rule describes the agent $A$ who sends a fresh message $m$ together with its MAC with the shared key $k$ to $B$.
Finally, the third rule describes $B$ who is expecting a message and a corresponding MAC with $k$ as input.
Note that the third rule can only be triggered if the input matches the premise, \ie if the input message is correctly MACed with $k$.
\[
\begin{array}{l}
Create\_Ltk : ~ [\factStyle{Fr}(k)] \rwr[] [\factStyle{!Ltk}(k)], \\
Send\_A : ~ [\factStyle{!Ltk}(k), \factStyle{Fr}(m)] \rwr[\factStyle{Sent}(m)] [\factStyle{Out}(\pair{m,\mac(m,k)})], \\
Receive\_B : ~ [\factStyle{!Ltk}(k), \factStyle{In}(\pair{m,\mac(m,k)})] \rwr[\factStyle{Received}(x)] [] \qed\\
\end{array}
\]
\end{example}
\looseness=-1
These rules yield a labeled transition system describing the possible protocol executions (see~\cite{tamarin-manual,schmidt2012automated} for  details on syntax and semantics).
\tamarin combines the protocol semantics with a Dolev-Yao~\cite{DY81} style attacker.
This attacker controls the entire network and can thereby intercept, delete, modify, delay, inject, and build new messages.
However, the attacker is limited by the cryptography: he cannot forge signatures or decrypt messages without knowing the key (black box  cryptography assumption).
He can nevertheless apply any function (e.g., hashing, XOR, encryption, pairing, \ldots) to messages he knows to compute new messages.

\subsection{Formalizing Security Goals in Tamarin}
\label{sec:formal:prop}

In \tamarin, security properties are specified in two different ways.
First, trace properties, such as secrecy or variants of authentication, are specified using formulas in a first-order logic with timepoints.
\begin{example}\label{ex:property}
Consider the multiset rewrite rules given in Example~\ref{ex:hashmsr}.
The following property specifies a form of non-injective agreement on the message,
\ie that any message received by $B$ was previously sent by $A$:
\\[0pt]\null\hfill$
    \forall i,m.
    \mathit{Received}(m)@i
    \Rightarrow ( \exists j .  \mathit{Sent}(m)@j \wedge j \lessdot i) .
$\hfill\null\\[2pt]
    Since the 5G AKA protocol features multiple roles and multiple instantiations thereof, agreement properties additionally require that the views of the two partners
  (who is playing which role, and what is the identity of the partner) actually match; see \Cref{ap:taxonomy}.
\end{example}
For each specified property, \tamarin checks that the property holds for all possible protocol executions, and all possible attacker behaviors.
To achieve this, \tamarin explores all possible executions in a
backward manner, searching for reachable attack states, which are counterexamples to the security properties.

Equivalence properties, such as unlinkability, are expressed by requiring that two instances of the protocol cannot be distinguished by the attacker.
Such properties are specified using \emph{diff}-terms
(which take two arguments), essentially defining two different instances of the protocol that only differ in some terms.
\tamarin then checks observational equivalence (see~\cite{tamarin-equiv}), i.e., it compares the two resulting systems and checks that the attacker cannot distinguish them for any protocol execution and any adversarial behaviors.

\looseness=-1
In fully automatic mode, \tamarin  either returns a proof that the property holds, or a counterexample/attack if the property is violated, or it may not terminate as the underlying problem is undecidable.
\tamarin can also be used in interactive mode, where the user can guide the proof search.
Moreover the user can supply heuristics called \emph{oracles} to guide the proof search in a sound way.
We heavily rely on heuristics in our analyses as they allow us to tame the protocol's complexity, as explained below.

\subsection{Modeling Choices}
\label{sec:formal:choices}
To better delimit the scope of our model and our analyses, we now describe some of our modeling choices.

  
\paragraph{Architecture}
We consider three roles (subscribers, \SNs, and \HNs) and reason
with respect to unboundedly many instances of each role.  As expected,
each subscriber credential is stored in at most one \HN.  We model
communication channels between these parties that provide security
properties as explained in \Cref{sec:spec:assumptions}.
Additionally, the messages exchanged are \emph{tagged} on the authenticated
channel between the \SNs and \HNs. This models the 
implicit assumption that the authenticated channel between an \SN and
an \HN role instance is protected from \emph{type flaw attacks}.

\paragraph{Modeling Cryptographic Messages}
We model and treat the subscribers' $\sqns$ as natural numbers (using
a standard encoding based on multisets~\cite{GroupProtocolsTamarin,tamarin-manual}).
We assume the attacker cannot follow \UEs from their creation so the
$\sqn$ is not known (see \Cref{sec:spec:assumptions}) at first, and we
thus start the sequence number with a random value. The freshness
check (\ie \textsf{(ii)} from \Cref{fig:5GAKA}) is faithfully modelled
as
a
natural number comparison.
Since the $\sqn$ may become out-of-sync during normal protocol execution,
we also consider an attacker who can arbitrarily increase
$\sqnUE$ (\UE does not allow decrease). Note that the attacker can
already increase $\sqnHN$ by repeatedly triggering authentication
material requests.  We fully model the re-synchronization mechanism
and let the \HNs update their $\sqnHN$ accordingly.  The concealment of
the $\sqn$, using Exclusive-OR (XOR), is faithfully modeled by relying
on the recent extension of \tamarin with equational theories including
XOR~\cite{tamarin-xor}.

\paragraph{Compromise Scenarios}
We model various compromise scenarios:
secret key reveals (of $\k$ or $\skHN$),
reveals of the $\supi$ or the initial value of $\sqn$, and
\SN compromises (\ie the attacker gains access to an authenticated channel with the \HNs).
This is needed mainly for two reasons.  First, the specification
itself considers some of those scenarios and still requires some
security guarantees to hold (cf.~the compromised \SNs
from \Cref{sec:spec:assumptions}). Second, this enables a comprehensive analysis to identify the minimal
assumptions required for a property to hold. For instance, if some
critical authentication property were violated when the attacker could
access the initial value of the $\sqn$, this would represent a potential
vulnerability in the protocol since the $\sqn$ is not a strong secret and
the search space of the $\sqn$ that the attacker needs to explore could be
further reduced by exploiting the meaning of this counter.

\paragraph{Implicit Authentication}
We equip the model with an optional key-confirmation roundtrip where
the subscribers and \SNs confirm their key $\kseaf$ by MACing
different constants. Our security analysis is then parametric in this
roundtrip, allowing us to derive which properties  hold without
key confirmation, and what is gained by including this key confirmation
step.

\paragraph{Simplifications Made}
\looseness=-1
As usual in the symbolic model, we omit message bit lengths. Some
key derivation functions also take the length of their arguments
to prevent type-flaw attacks. This is covered in our
model as such length-based misinterpretation cannot happen.
The protocols under study feature some sub-messages that are publicly
known constants, for example, fixed strings like AMF, ABBA, or \cst{MAC\_Failure}.  We
mostly omit such sub-messages, unless they are useful as tags.
We do not model the optional, non-nor\-ma\-tive protection against
wrapping around the $\sqn$\linebreak[4] \spec{TS\,33.102}{C}. Note that this is in line
with our modeling of the $\sqn$ as a natural number for which no wrapping
can occur.
The 5G AKA protocol establishes a session key, to
which a key identifier is associated (the key set identifier ngKSI).
Such identifiers are needed for subsequent procedures only and do not
interact with the authentication methods and hence we omit them.
An \SN may create a pseudonym, called 5G-GUTI, associated with the
$\supi$ of a subscriber who is visiting this \SN, in order to
recognize this subscriber in a subsequent session. We omit this
optional mechanism.
Authentication tokens do not expire in our model as is usual in
symbolic models. However, since such mechanisms are never clearly specified in normative documents, 
we emphasize that critical security properties should not rely on them.

\subsection{Tamarin Models of 5G AKA}
\label{sec:formal:models}
We have built a \tamarin model for the 5G AKA authentication method which enables automated security analyses.
Our models and associated documentation are available online~\cite{tam-models}, and use \tamarin v1.4.0~\cite{tam-tool-release}, which includes XOR support.

Writing a formal model of such a substantial real-world protocol is challenging. However, the real difficulty is
doing this in a way that enables effective reasoning about the models, \ie is amenable to automation.
We now describe this modeling as well as the proof strategies we  developed,
and argue why this can serve as a basis for future analyses of protocols in the AKA family.

\subsubsection{Challenges}
The 5G AKA protocol uses a combination of features that make reasoning about these models highly complex.
First, 5G AKA is a stateful protocol, \ie it relies on internal states (the $\sqns$) that are persistent across sessions and that are mutable. In the symbolic model, the set of values these states can take --- all natural numbers --- is unbounded.
This feature alone excludes most verification tools.
Verifiers for a bounded number of sessions are not a viable choice, simply due to the size of a single session.
Moreover, the  sequence numbers are not only internal counters,
they are also used for comparison on input. This requires the ability to compare
two values (see \Cref{sec:formal:choices}) in the chosen
representation of natural numbers. This is demanding in terms of proof
efforts: to the best of our knowledge, this is the first time a
complete, real-world protocol relying on natural numbers and
comparisons is analyzed with an automated formal verifier in the unbounded setting. Previous examples are
limited to the case of just an internal counter for a
TPM~\cite{thesis-simon} or small examples, like
simplified Yubikey~\cite{yubikey}.

Second, 5G AKA heavily relies on XOR to conceal the value of $\sqns$. 
Reasoning about XOR in the symbolic model is  challenging and
its integration in \tamarin is recent~\cite{tamarin-xor}.  Intuitively,
this is because of the intricate algebraic properties of XOR
(\ie associativity,
commutativity, cancellation, and neutral element).
This
considerably increase 
the search space  when proving properties.
Again, in the symbolic model, we are not aware of any formal analysis of such a large-scale real-world
protocol featuring XOR.

Finally, the state-machine of the 5G AKA protocol is large and complex.
Role instantiations can be in 14 different states.
Evolution between those states includes numerous
loops, notably because of the persistent and mutable states' $\sqns$,
\eg sessions can be repeated while using a given $\sqn$.

\subsubsection{Proof Strategies}
The way $\sqns$ are updated on the subscribers' and \HN's sides, in particular with the re-synchronization procedure, induces complex state-changes that must be tackled by our proof strategies.
Manual proofs are not feasible due to the size of the search space one would have to explore.
In contrast, \tamarin's fully automatic mode fails to prove relevant security properties and even extremely weak properties such as the full executability of the protocols.
Our work straddles this divide: we developed a proof structure based on intermediate
lemmas (called \emph{helping lemmas}) as well as proof strategies for proving these lemmas and the security properties.
Proof strategies are implemented through \emph{oracles} that offer a light-weight tactic language, implemented in Python,
to guide the proof search in \tamarin.

The key helping lemmas we prove state that the $\sqn$ associated to a subscriber stored on his side (respectively on its \HN's side) is strictly increasing (respectively monotonically increasing).
Thanks to our chosen modeling of the states' $\sqns$ as multisets and the comparisons of $\sqns$ based on pattern-matching, we were able to prove the aforementioned lemmas by induction with a simple, general proof strategy.
The security properties, however, require dedicated and involved proof strategies (${\sim}$1000 LoC of Python).
The effort of writing such generic proof strategies represents several person-months.

\subsubsection{Our Models}
Based on our modeling choices, we built a complete model of 5G AKA (preceded by the initialization protocol) that is amenable to automation. We model fully parametric compromise scenarios that enables one to easily choose what kind of reveals or compromises are considered when proving  properties.
We also implement the key confirmation roundtrip in a modular way: one can consider authentication properties
after  this roundtrip or without.
The protocol model itself consists of roughly 500 LoC.

Our model includes all the necessary lemmas: helping lemmas, sanity-check lemmas, and the lemmas that check the relevant security properties against the 5G AKA protocol. Since we aim at identifying the minimal assumptions required for the stated properties to hold, we prove several lemmas for each security property.
First, we state a lemma showing that the property holds under a certain set of assumptions.
Second, we show the minimality of this set of assumptions. We do this by disproving all versions of the previous lemma where the set of assumptions is reduced by just one assumption.
This requires 124 different lemmas and ca.~1000 LoC.
\tamarin needs ca.~$5$ hours to automatically establish all the proofs and find all the attacks.

Our model of 5G AKA is general in that it can be used to model all other protocols from the AKA family
requiring only localized modifications in the model.
Part of the model (creation or role instantiations, reveal and compromise modelings, \etc) would not change, but the roughly 300 LoC defining the main flow of the protocol would have to be adapted. The size of this change depends on how  different  the chosen protocol is to 5G AKA.
We expect our oracle to be still valid, at least after minor modifications to the model.
Furthermore, given that our analysis is fully automatic (thanks to our proof strategies), our model can be easily kept
up-to-date as the standard further evolves and any change in terms of provided security guarantees can then be automatically identified by the tool.

\definecolor{darkgreen}{rgb}{0.0, 0.2, 0.13}
\definecolor{darkred}{rgb}{0.55, 0.0, 0.0}
\definecolor{cadmiumgreen}{rgb}{0.0, 0.42, 0.24}
\newcommand{\bad}[1]{\textcolor{darkred}{#1}}
\newcommand{\good}[1]{\textcolor{cadmiumgreen}{#1}}
\newcommand{\Res}[1]{{\small{#1}}}
\newcommand{\ResTODO}{\Res{\phantom{K}?\phantom{K}}}
\newcommand{\Asupi}{\Res{$\supi$}}
\newcommand{\Ak}{\Res{$\k$}}
\newcommand{\Asqn}{\Res{$\sqn$}}
\newcommand{\AskHN}{\Res{$\skHN$}}
\newcommand{\Achannel}{\Res{\good{ch}}}
\newcommand{\Akc}{\Res{\good{k-c}}}
\newcommand{\Anot}[1]{\Res{\good{$\lnot$#1}}}
\newcommand{\Arev}[1]{\Res{\bad{#1}}}
\newcommand{\Anone}{\Res{\good{$\emptyset$}}}
\newcommand{\Aand}{\Res{\bad{$\land$}}}
\newcommand{\Aor}{\Res{\good{$\lor$}}}
\newcommand{\Aattack}{\Res{\attack}}
\newcommand{\Ano}{$\times$}
\newcommand{\Ared}{wa}
\newcommand{\Arevel}{${-}$}
\newcommand{\SPEC}[1]{[#1]}


\newcommand{\ra}[1]{\renewcommand{\arraystretch}{#1}}
\renewcommand{\sp}{\phantom{a}}
\newdimen\POV
\newdimen\PART
\newdimen\DIR
\POV=0.15em
\PART=0.1em
\DIR=0.03em

\begin{table*}[th]
  \centering
  \ra{0.7}
  \begin{tabular}{@{}l
    cccc
    c
    cccc
    c
    cccc
    @{}}
    \toprule \midrule
    Point of view
    & \multicolumn{4}{c}{\UE} & \sp
    & \multicolumn{4}{c}{\SN} & \sp
    & \multicolumn{4}{c}{\HN}\\
    \cmidrule[\POV](r){2-5}
    \cmidrule[\POV](r){7-10}
    \cmidrule[\POV](r){12-15}
    Partner
    & \multicolumn{2}{c}{\SN}
    & \multicolumn{2}{c}{\HN} & \sp
    & \multicolumn{2}{c}{\UE}
    & \multicolumn{2}{c}{\HN} & \sp
    & \multicolumn{2}{c}{\UE}
    & \multicolumn{2}{c}{\SN}
    \\
    \cmidrule[\PART](r){2-3}
    \cmidrule[\PART](r){4-5}
    \cmidrule[\PART](r){7-8}
    \cmidrule[\PART](r){9-10}
    \cmidrule[\PART](r){12-13}
    \cmidrule[\PART](r){14-15}
    Agreement
    & NI & I
    & NI & I & \sp
    & NI & I
    & NI & I & \sp
    & NI & I
    & NI & I\\
    \cmidrule[\DIR](r){2-2}
    \cmidrule[\DIR](r){3-3}
    \cmidrule[\DIR](r){4-4}
    \cmidrule[\DIR](r){5-5}
    \cmidrule[\DIR](r){7-7}
    \cmidrule[\DIR](r){8-8}
    \cmidrule[\DIR](r){9-9}
    \cmidrule[\DIR](r){10-10}
    \cmidrule[\DIR](r){12-12}
    \cmidrule[\DIR](r){13-13}
    \cmidrule[\DIR](r){14-14}
    \cmidrule[\DIR](r){15-15}
    on $\kseaf$
    & \Aattack & \Aattack & \Anot{\Ak}\Aand\Akc & \Anot{\Ak}\Aand\Akc & \sp   
    & \Aattack & \Aattack & \Anot{\Achannel} & \Anot{\Ak}\Aand\Anot{\Achannel} & \sp 
    & \Anot{\Ak} & \Anot{\Ak} & \Anot{\Achannel} & \Anot{\Achannel} \\    
    on $\supi$
    & \Ared & \Ano & \Ared & \Ano & \sp 
    & \Ared & \Ano & \SPEC{\Anot{\Achannel}} & \Ano & \sp 
    & \Ared & \Ano & \Ano & \Ano \\    
    on $\SNname$
    & \Ared & \Ano & \SPEC{\Anot{\Ak}\Aand\Akc} & \Ano & \sp 
    & \Ared & \Ano & \Ared & \Ano & \sp 
    & \SPEC{\Anot{\Ak}} & \Ano & \Ared & \Ano \\    
    \addlinespace
    \cmidrule[\PART](r){2-3}
    \cmidrule[\PART](r){4-5}
    \cmidrule[\PART](r){7-8}
    \cmidrule[\PART](r){9-10}
    \cmidrule[\PART](r){12-13}
    \cmidrule[\PART](r){14-15}
    Weak agreement
    & \multicolumn{2}{c}{\SPEC{\Aattack}}       
    & \multicolumn{2}{c}{\Anot{\Ak}} & \sp 
    & \multicolumn{2}{c}{\SPEC{\Anot{\Ak}\Aand\Anot{\Achannel}}}       
    & \multicolumn{2}{c}{\Anot{\Achannel}} & \sp 
    & \multicolumn{2}{c}{\Anot{\Ak}}       
    & \multicolumn{2}{c}{\Anot{\Achannel}}       
    \\

    \midrule \bottomrule
  \end{tabular}
  \caption{Minimal assumptions required for 5G AKA to achieve authentication properties.
    We denote subscribers by \UE, non-injective by NI, and
    injective by I.
    Assumptions are expressed in terms of
    \good{forbidden reveals} (\eg~\Anot{\Ak}, meaning the property only holds when \Ak~is not revealed).
    We also indicate whether a key confirmation phase is needed
    with~\good{\Akc} while
    \Anot{\Achannel}~denotes an uncompromised channel between \SN and \HN.
    When not otherwise  specified, the worst-case scenario is considered;
    that is $\k$,${\normalfont \sqn}$,${\normalfont \supi}$,$\mathit{sk}_{\mathsf{HN}}$, and the channel between \SN and \HN are compromised and
    the key confirmation phase is skipped.
    \Ano: the property is violated {\em by definition} (\eg because ${\normalfont \supi}$ is constant).
    \Ared: the property coincides with weak agreement and requires same assumptions.
    The explicit goals given in the specification are denoted by \SPEC{$\cdot$} around them.
  }
  \label{fig:results:auth}
\end{table*}

\begin{table}[t]
  \centering
  \ra{0.7}
  \begin{tabular}{@{}l
    ccc               
    @{}}
    \toprule
    Point of view & \UE & \SN & \HN \\
    \midrule
    $\kseaf$ & \Anot{\Ak}\Aand\Anot{\Achannel} & \Anot{\Ak}\Aand\Anot{\Achannel} & \Anot{\Ak}\Aand\Anot{\Achannel} \\
    $\textsf{PFS}(\kseaf)$ & \Aattack & \Aattack & \Aattack \\
    $\supi$  & \Anot{\AskHN}\Aand\Anot{\Achannel}$^*$ & \Arevel & \Anot{\AskHN}\Aand\Anot{\Achannel}$^*$ \\
    $\k$     & \Anone & \Anone & \Anone \\    
    \bottomrule
  \end{tabular}
  \caption{Minimal assumptions for 5G AKA to achieve secrecy properties.
    We omit the assumption that data that is supposed to be secret is not revealed.
    See \Cref{fig:results:auth} for the legend.
    The symbol $^*$ indicates that there is no dishonest \SNs at all and the underlying property is always violated otherwise.
    \textsf{PFS}($\cdot$): perfect-forward secrecy.
    \Arevel: property not relevant.
  }
  \label{fig:results:secrecy}
\end{table}

  

\renewcommand{\sp}{}
\begin{table}[t]
  \centering
  \ra{0.7}
  \begin{tabular}{@{}l
    cc
    c
    cc
    @{}}
    \toprule \midrule
    P.o.V.
    & \multicolumn{2}{c}{\UE} & \sp
    & \multicolumn{2}{c}{\SN} \\
    \cmidrule[\POV](r){2-3}
    \cmidrule[\POV](r){5-6}
    Partner
    & \multicolumn{2}{c}{\SN} & \sp
    & \multicolumn{2}{c}{\UE}
    \\
    \cmidrule[\PART](r){2-3}
    \cmidrule[\PART](r){5-6}
    Agre.
    & NI & I & \sp
    & NI & I\\
    \cmidrule[\DIR](r){2-2}
    \cmidrule[\DIR](r){3-3}
    \cmidrule[\DIR](r){5-5}
    \cmidrule[\DIR](r){6-6}
    on $\kseaf$
    & \Anot{\Ak}\Aand\Akc\Aand\Anot{\Achannel} & \Anot{\Ak}\Aand\Akc\Aand\Anot{\Achannel} & \sp 
    & \Anot{\Ak}\Aand\Anot{\Achannel} & \Anot{\Ak}\Aand\Anot{\Achannel} \\    
    \cmidrule[\PART](r){2-3}
    \cmidrule[\PART](r){5-6}
    Weak agre.
    & \multicolumn{2}{c}{\SPEC{\Anot{\Ak}\Aand\Akc\Aand\Anot{\Achannel}}} & \sp 
    & \multicolumn{2}{c}{\SPEC{\Anot{\Ak}\Aand\Anot{\Achannel}}} \\    
    \midrule \bottomrule
  \end{tabular}
  \caption{Minimal assumptions required for 5G AKA to achieve authentication properties
    between \UEs and \SNs, assuming that the channel between \HNs and \SNs is binding.
    Agreements on $\supi$ and $\SNname$ are not impacted.
  }
  \label{fig:results:auth:asssumption}
\end{table}


\section{Security Analysis}
\label{sec:analysis}

\subsection{Results}
\label{sec:analysis:res}
We present the results of our comprehensive analysis of the 5G AKA protocol.
We emphasize that we \emph{automatically} analyze the \emph{formal security guarantees} that the protocol provides
for an \emph{unbounded number} of sessions executed by honest and compromised subscribers, \SNs, and \HNs
when used \emph{in combination} with the initiation protocol.
Thus, our analysis accounts for all potential unintended interactions an attacker could exploit between these sub-protocols
run by all possible instantiations of the three roles we consider.

We depict the outcome of our analysis of authentication properties
in \Cref{fig:results:auth}.
For each pair of parties,
we present the minimal assumptions required to achieve
authentication properties:
\ie weak agreement, non-injective agreement,
and injective agreement.
We only consider agreement on relevant data; \ie $\kseaf$, $SNname$, and the $\supi$ (recall that the $\supi$ already contains $\idHN$).
The assumptions are minimal in  that strengthening
the attacker's capabilities
in any direction violates the property.
The symbol \Aattack~denotes that the property is violated for the
weakest threat model where all participants are honest, none of the compromise scenarios is considered,
and key confirmation is systematically enforced.
Similarly, we present our results concerning secrecy properties in \Cref{fig:results:secrecy}.

We only check for 2-party authentication properties, which expresses well the security goals of 5G AKA.
Note however that we obtain a form of 3-party agreement property (where all 3 parties' views coincide) as a corollary
of three 2-party agreement properties. This is because we check for 
strong 2-party agreement properties on several data points and identifiers simultaneously.

\subsection{Discussion}
\label{sec:analysis:discussion}
\Cref{fig:results:auth} clearly shows the extent that the 5G standard underspecifies authentication
requirements (recall that $[\cdot]$ denotes explicit goals); see \Cref{sec:prop:under}. 
We also indicate a number of properties that are violated even in the best-case scenario (\Aattack).
We discuss why in \Cref{sec:analysis:discussion:missing}.
Afterwards, we explain and critique the use of key confirmation in \Cref{sec:analysis:discussion:implicit}.
We discuss privacy properties in \Cref{sec:analysis:discussion:privacy}.
Finally, our results concerning secrecy properties are as expected and are
not discussed further. Also, perfect forward secrecy of $\kseaf$ is violated as expected.

\subsubsection{Missing Security Assumption}
\label{sec:analysis:discussion:missing}
The 5G AKA protocol fails to meet several security goals that are explicitly
required as well as other critical security properties.
This is still true under the assumptions specified in the standard,
even after a successful key-confirmation phase (see \Aattack~in \Cref{fig:results:auth}).
More specifically, the agreement properties on $\kseaf$ between the subscribers and \SNs are
violated. So is weak agreement from the subscribers towards the \SNs.
This is caused by the lack of a binding assumption on the channel between \SNs and \HNs and because
the $\supi$ is sent to the \SN in a different message than the message containing $\kseaf$, which is sent earlier.
Therefore, as soon as a pair of an \SN and an \HN runs two sessions concurrently,
there is no assurance that the $\supi$ the \SN receives at the end of the protocol actually corresponds to
the $\kseaf$ it has received earlier (it could correspond to another concurrent session).
As a consequence, an \SN may associate the session key $\kseaf$ to the correct subscriber (a necessary condition for the key confirmation to be successful),
but to the wrong $\supi$, violating the aforementioned properties.
In practice, this could allow an attacker to make the \HN bill someone else (\ie with a different $\supi$) for
services he consumes from an \SN (\ie encrypted with $\kseaf$).
Thus, the binding property for the channel between the \SNs and \HNs appears to be a critical security assumption, and should be explicitly mentioned in the standard.
This weakness has been introduced in the version \texttt{v0.8.0} of the standard (published in March 2018). In the previous version (\texttt{v0.7.1}), the $\supi$ was sent
by the \HN to the \SN together with the challenge\footnote{To the best of our knowledge, the rationale behind the new version is to let \HN wait for the proof of the subscriber's recent aliveness before disclosing the corresponding $\supi$ to \SNs that may be malicious or dishonest.},
thus preventing the aforementioned attack and making the binding assumption unnecessary. However, the final version of the standard
requires this additional assumption.
A similar looking issue, but between two parts of the \HN has been previously observed by~\cite{casmartin-binding}, but it is an entirely different concern than the one we describe.

\Cref{fig:results:auth:asssumption} depicts additional security properties the 5G AKA protocol provides when the channel between the \SNs and \HNs
is assumed to be binding.
Under this  assumption, the previously violated properties are now satisfied under reasonable threat models. We only show results for \UEs and \SNs to show how their guarantees change.


\subsubsection{Implicit Authentication}
\label{sec:analysis:discussion:implicit}
A successful key-confirmation\linebreak[4]
roundtrip is required to obtain crucial security guarantees.
More precisely, this roundtrip is required for all agreement properties from the subscribers' point of view
except weak agreement towards the \HNs. Indeed, an attacker can impersonate an \SN towards a subscriber but is unable
to learn the $\kseaf$ key the subscriber has computed.
\quoteTS{5.1.2}{\size
  The meaning of `implicit key authentication' here is that authentication is provided through the successful
  use of keys resulting from authentication and key agreement in subsequent procedures.
}
The 5G standard only requires \emph{implicit} authentication properties for the subscribers.
However, the standard neither specifies that subscribers must wait for this key confirmation to be successful before continuing
nor does it specify this additional roundtrip \emph{as part} of the authentication method.
As a consequence, the standard makes a choice that we consider risky: it postpones the handling and the verification of the additional
key confirmation roundtrip to \emph{all possible subsequent procedures} (\eg the NAS security mode command procedure \specTS{6.7.2}).
The standard fails to specify a standalone authentication
protocol that provides a reasonable set of security guarantees since some critical properties are provided only when the protocol is used
in specific, appropriate contexts.

More importantly, since the standard makes the overall security of the authentication rest on subsequent procedures, it is very challenging,
and out of the scope of the present paper, to assess if all currently specified subsequent procedures (as well as future ones that may be added) either correctly mandate the use of this key confirmation roundtrip
or do not require authentication properties from the subscribers' point of view towards the \SNs.
We believe that there are at least two potential use cases where the above weakness represents a vulnerability.
First, the standard specifies that \SNs can initiate \emph{key change on-the-fly} \specTS{6.9.4.1} as well as switch security contexts \specTS{6.8}, including keys, parameters, \etc.
This raises the question whether a malicious \SN or a fake base station could not \emph{fully} impersonate a genuine \SN towards the subscribers
by changing the session key immediately after 5G AKA.
Second, in a scenario where subscribers use the presence of \SNs for geo-localization
or for making sensitive decisions (related to \eg emergency calls),
an active attacker could impersonate an \SN since the (mismatched) $\kseaf$ key may not be needed or used.

Finally, 
the key confirmation roundtrip is not the only option to achieve the aforementioned missing security guarantees.
We provide and discuss in \Cref{sec:analysis:rec:keyConf} two alternative solutions that fix this issue
while reducing neccessary communications.


\subsubsection{On Privacy}
\label{sec:analysis:discussion:privacy}
As mentioned in \Cref{sec:spec:prop:privacy}, the 5G standard aims to protect
privacy only against passive attackers.
5G AKA provides an identifier hiding mechanism and sends the $\supi$ only 
in a randomized public key encryption (the Subscription Concealed Identifier, $\suci$).
We show with \tamarin that the $\supi$ indeed remains confidential, even
against active attackers (see \Cref{fig:results:secrecy}) and hence also
against passive attackers. 
5G AKA thus defeats previous active
IMSI-catcher attacks~\cite{mobileSoK17}, which relied on the subscribers sending the IMSI (matching $\supi$ in 5G) in the clear.
We also have modelled a weak, passive attacker and have automatically proven that he cannot
trace subscribers.

We believe that active attackers are realistic threats for most use cases.
Moreover, since privacy is a real concern to the 3GPP, 5G AKA should protect subscribers' privacy also against active attackers.
Unfortunately, we have found that this is not the case as the 5G AKA protocol suffers from a traceability attack.

Using \tamarin (see our model~\cite{tam-models}), we automatically find the following attack in 5G AKA.
In this attack, the attacker observes one 5G AKA authentication session and later
replays the \SN's message to some subscriber. From the subscriber's answer (MAC failure or Synchronization failure), the attacker
can distinguish between the subscriber observed earlier (in case of Synchronization failure) and a different
subscriber (in case of MAC failure).
This attack can be exploited to track subscribers over time.
%
A variant of this attack was first
described in~\cite{arapinis2012new} for the AKA protocol as used in 3G.
%


\subsection{Recommendations}
\label{sec:analysis:rec}
Throughout the paper, we have highlighted weaknesses in the standard and suggested improvements and refinements.
We now summarize some of them and propose more precise, provably
secure fixes as a replacement for the key confirmation and the binding
channel 
assumptions.
Again, we emphasize the critical role  played here by our formal interpretation of the standard and our formal analysis of the described 5G AKA protocol.

\subsubsection{Explicit Requirements}
As shown in \Cref{fig:results:auth} and discussed in \Cref{sec:prop:under}, the standard underspecifies security requirements for the 5G AKA protocol. We suggest that the standard explicitly requires the missing intended security properties.
In particular, it should be clear that 5G AKA aims at achieving \prop{injective agreement on $\kseaf$ between the subscribers and the \SNs} which is central to the protocol's purpose.
The subscribers should obtain \prop{injective agreement on $\kseaf$ with the \HNs}; they are thereby assured the \HNs~\emph{recently} authorized
this session, since $\kseaf$ is derived from the random $\r$.
Finally, the \HNs should have \prop{injective agreement on $\kseaf$ with the subscribers}, obtaining recent aliveness as a consequence.

\subsubsection{Binding Channel}
As discussed in \Cref{sec:analysis:discussion:missing}, a recent update in the standard introduced attacks under the given security assumptions. There are two solutions to fix this: either the standard  explicitly states an additional security assumption (\ie the channel between the \SNs and \HNs must be binding),
or alternatively the 5G AKA protocol is fixed (without the need for a new assumption) using the following minor modification:
$\tuple{\supi,\suci}$ is sent instead of $\supi$ in the final message from \HN to \SN.
However, it is our understanding that the binding assumption is a property that is required for other reasons anyway, such as reliability.

\subsubsection{On the Key Confirmation}
\label{sec:analysis:rec:keyConf}
We already have discussed the danger of missing key confirmation in 5G AKA in \Cref{sec:analysis:discussion:implicit}.
We now propose two simple modifications to the protocol that would make key confirmation redundant and unnecessary, therefore reducing the number of roundtrips that are needed to achieve intended security guarantees.
Before explaining our fix, note that the key confirmation was necessary in the first place because the \HNs never commit to a specific \SNname when computing the challenge $\r,\AUTN$. Only the key $\kseaf$ is bound to $\SNname$, but the challenge itself is not.

Our first fix consists of binding $\AUTN$ to $\SNname$ so that subscribers directly have the proof the \HN has committed to a specific \SNname,
without even using $\kseaf$. Formally, $\AUTN$ currently refers to $\tuple{\sqnHN \xor \m{AK}, \m{MAC}}$
where $\m{MAC} = \one(\k,\tuple{\sqnHN, \r})$. In our fix, $\m{MAC}$ is replaced by
$
\one(\k,\tuple{\sqnHN, \r, \SNname})$.
Therefore, the subscribers can verify the authenticity of the challenge that commits to a specific \SNname.
We have formally verified~\cite{tam-models} that a key-confirmation roundtrip is no longer necessary with this fix.

Our second, alternative, fix consists in replacing the full key-confirmation roundtrip by an unidirectional key confirmation from the \SN only. More precisely, we could add (any) message MACed with a key derived from $\kseaf$, sent by the \SNs to the subscribers, at the very end of the protocol.
We have proven with \tamarin that no further guarantees are provided by a full key confirmation, compared
to our (less costly) unidirectional key confirmation.

\subsubsection{On Privacy}
\label{sec:analysis:rec:privacy}
We recall that the functions $\one$ and $\ones$ are not explicitly required to protect the
confidentiality of their inputs (see \Cref{sec:spec:assumptions:primitives}).
This is however necessary for privacy as these MAC functions take $\sqn$
as input, among others. If these functions were not confidentiality-preserving, a passive attacker could learn the
subscribers' $\sqns$ and perform location attacks~\cite{mobileSoK17}
by tracking nearby $\sqns$ over time or perform
activity monitoring attacks~\cite{BH-BH17}.

We also recommend for the standard to explicitly aim at protecting privacy
against active attackers and take steps in this direction.
Unfortunately, this would involve
significant
modifications to the protocol since at least
the failure reasons (MAC\slash Synchronization failure) must be hidden from the attacker~\cite{arapinis2012new,fouque2016achieving}
and the $\sqn$ concealment mechanism should be strengthened against active attackers~\cite{BH-BH17}, possibly by using proper
encryption or using an anonymity key \m{AK} based on subscriber-generated randomness.
We leave a complete evaluation of possible solutions for
future work and we expect our model to be valuable for this process.

\subsubsection{Redundancies}
\looseness=-1
A close look at the cryptographic messages (see detailed list in \Cref{ap:index})
and their purposes shows many redundancies.
For instance, in \m{RES\s}, the proof of possession of $\k$ is in \m{CK}, \m{IK}, and \m{RES}.
$\r$ appears to be redundant as well. 
Similarly, $\SNname$ is redundant in the key derivation of $\kseaf$. 
Legacy reasons may explain these
redundancies, but 
these design choices could be questioned and the protocol simplified.

\subsubsection{On the Role of $\sqn$}
The purpose of the $\sqn$ counters is to provide replay protection for
the subscribers. This mechanism  was introduced in 3G, when the USIM
was incapable of generating good randomness. This is no longer the
case in 5G, where USIMs can perform randomized asymmetric encryption 
(required to compute $\suci$ from $\supi$). Therefore, authentication protocols should be rethought and more standard challenge-response mechanisms could be used to replace the $\sqn$ counters.
This would benefit the current authentication methods, which can suffer from de-synchronization and must keep the privacy sensitive $\sqns$ up-to-date and sometimes fail to protect them against attackers (see \Cref{sec:analysis:rec:privacy}).

\subsubsection{On the Benefits of Formal Methods}
As argued throughout this section, the standard could be simplified and improved in various directions.
We recall that formal models, such as our model of the 5G AKA protocol, have proven to be extremely valuable to quickly assess the security of such
modifications and simplifications.
Our model can serve as a basis to accompany the standard's future evolution
and provides a tool for 
quickly evaluating the security
of modification proposals.

\section{Conclusion}
\label{sec:conclusion}

We have formally analyzed  one of the two authentication
methods in 5G, the one which enhances the previous variant currently 
used in 4G.  This included a detailed analysis of the standard to
identify all assumptions and security goals, a formal model of the
protocol and security goals as specified in the standard, the automated
security analysis using the \tamarin prover, and a detailed discussion
of our findings.  Our models are substantially more detailed than those
of previous work and account for details of
the state machine, counters, the re-synchronization procedures, and the
XOR operations.

While analyzing the standard we discovered that security goals and
assumptions are underspecified or missing, including
central goals like agreement on the session key.  Moreover, our
analysis in \tamarin shows that some properties are violated
without further assumptions. A striking example of this is agreement
properties on the session key.  We also critique the standard's choice
of implicit authentication and the lack of key confirmation as this
introduces weaknesses if the protocol is used in ways other than intended.
Finally, our privacy analysis shows that the 5G version of AKA still
fails to ensure unlinkability against an active attacker; this
scenario is, in our opinion, completely realistic.

As future work, we plan to analyze  other variants of the
AKA protocol, notably those used in 3G and 4G networks, to see which security guarantees they provide compared to 5G AKA.
We will also follow the future development of the 5G standard as our analysis can serve as the basis for improving the protocol's design, in particular to evaluate ideas and avoid regressions.
For example, we identified one weakness that was introduced in a recent update (from \texttt{v0.7.1} to \texttt{v0.8.0}).
This is a major benefit of tool-based  analysis of protocol
design: once the model is constructed, one can quickly test changes and
evaluate different design options.

\begin{acks}
  We are grateful for the support from the
  \grantsponsor{SPOOC}{EUs Horizon 2020 research and innovation program}{https://ec.europa.eu/programmes/horizon2020/}
  under ERC Grant No.:~\grantnum{SPOOC}{645865-SPOOC}.
  %
  The authors also thank Huawei Singapore Research Center
  for their support for parts of this research.
  %
  %
\end{acks}

\bibliographystyle{ACM-Reference-Format}
\bibliography{refs}

\appendix
\section{Notations and Acronyms}
\label{ap:index}

We list all the acronyms we introduced throughout the paper in \Cref{tab:acronyms} and give a correspondence table between
our simplified terminology and the 3GPP terminology in \Cref{tab:correspondence}.

We describe the cryptographic messages format in \Cref{tab:notations} (abstracting away AMF, other constants and sub-message lengths).

\begin{table*}[ht]
  \centering
  \begin{tabular}[]{|c|l|l|l|}
    \hline
    Acronym & Full name & Reference & Description (if needed) \\\hline\hline
    \m{AK}     & Anonymity Key & \Cref{sec:spec:protos} & \\
    \m{AMF} &  Authentication Management Field & \Cref{ap:index} & \\
    AMF$^{(*)}$  & Access and Mobility Management Function & \Cref{ap:index} & \\                    
    ARPF & Authentication credential Repository and Processing Function & \Cref{ap:index} & \\
    AUSF & Authentication Server Function & \Cref{ap:index} & \\
    EAP & Extensible Authentication Protocol & \Cref{sec:spec:protos} & \\
    \imsi & International Mobile Subscriber Identity & \Cref{sec:spec:archi} & Uniquely identify subscribers \\
    gNB & NR Node B & \Cref{ap:index} & new generation base station \\
    \m{GUTI} & Globally Unique Temporary UE Identity & \Cref{sec:formal:choices} & \\
    ME & Mobile Equipment & \Cref{sec:spec:archi} & \\
    MNC & Mobile Country Code & \Cref{ap:index} & Uniquely identify \HNs' countries \\
    MNC & Mobile Network Code & \Cref{ap:index} & Uniquely identity \HNs in a country\\
    \m{RAND} ($\r$) & Random Challenge & \Cref{sec:spec:protos} & \\
    \m{RES} & RESponse & \Cref{ap:index} & \\
    SEAF & SEcurity Anchor Function & \Cref{ap:index} & \\
    \suci & Subscription Concealed Identifier & \Cref{sec:spec:protos} & \\
    \supi & Subscription Permanent Identifier & \Cref{sec:spec:archi} & Uniquely identify a subscriber and its \HN \\
    \sqn  & SeQuence Number & \Cref{sec:spec:archi} & \\
    UDM & Unified Data Management & \Cref{ap:index} & \\
    USIM & Universal Subscriber Identity Module & \Cref{ap:index} & \\
    \m{XRES} & eXpected RESponse & \Cref{ap:index} & \\
    \m{SIDF} & Subscription Identifier De-concealing Function & \Cref{ap:index} & \\
    \m{SUPI} & Subscription Concealed Identifier & \Cref{sec:spec:archi} & \\
    \m{SUCI} & Subscription Permanent Identifier & \Cref{sec:spec:protos} & Randomized encryption of \m{SUPI} \\
    \m{SNid} & \SN identity & \Cref{ap:index} &  Uniquely identify \SNs \\
    \hline
  \end{tabular}
  \vspace{10pt}
  \caption{Acronyms and abbreviations (mostly from \specTS{3.2})}
  \label{tab:acronyms}
\end{table*}

\begin{table*}[ht]
  \centering
  \begin{tabular}[]{|r|l|}
    \hline
    Our notion & Correspondent notion in \texttt{TS33.501} \\
    \hline\hline
    Serving Network & Combination of SEAF, AMF$^{(*)}$ and gNB \\
    Home Network & Combination of AUSF, ARPF, UDM and SIDF \\
    \hline
  \end{tabular}
  \vspace{10pt}
  \caption{Correspondence between our terminology with the one of \texttt{TS33.501}}
  \label{tab:correspondence}
\end{table*}

\begin{table*}[ht]
  \centering
  \begin{tabular}{|r|l|l|l|}
    \hline
    Message Name & Content & Internal Ref. & Specification \\\hline\hline
    \m{SUPI}   & $\tuple{\imsi,\m{MMC},\m{MNC}}$ & \Cref{sec:spec:archi} & \spec{TS23.501}{5.9.2} \\
    \m{SUCI}   & $\tuple{\aenc(\tuple{\supi, \r_s}, \pkHN), \m{MMC},\m{MNC}}$ & \Cref{sec:spec:protos} & \specTS{C.3} \\
    \SNname   & $\tuple{\cst{5G},\cst{:},\SNid}$ & \Cref{sec:spec:archi} & \specTS{6.1.1.4} \\
    \hline
    \m{MAC} & $\one(\k,\tuple{\sqnHN, \r})$ & \Cref{sec:spec:protos} & \spec{TS\,133.102}{6.3.2}\\
    \m{AK}   & $\five(\k,\r)$ & \Cref{sec:spec:protos} & \spec{TS\,133.102}{6.3.2} \\
    \m{AUTN} & $\tuple{\sqnHN \xor \m{AK}, \m{MAC}}$ & \Cref{sec:spec:protos} & \specTS{6.1.3}\\
    \m{RES} & $\two(\k,\r)$ & \Cref{sec:spec:protos} & \spec{TS\,133.102}{6.3.2} \\
    \m{CK} & $\three(\k,\r)$ & \Cref{ap:index} & \spec{TS\,133.102}{6.3.2} \\
    \m{IK} & $\four(\k,\r)$ & \Cref{ap:index} & \spec{TS\,133.102}{6.3.2} \\
    \m{RES\s} & $\KDF(\tuple{\m{CK},IK},
           \tuple{\SNname,\r,\m{RES}})$ & \Cref{sec:spec:protos} & \specTS{A.4}\\
    $\CHALLENGE(\k,\r,\SNname)$ & \m{RES\s} & \Cref{sec:spec:protos} & None \\
    \m{HXRES\s} & $\sha(\tuple{\r,\m{RES\s}})$ & \Cref{sec:spec:protos} & \specTS{A.5} \\
    \hline
    \m{MACS} & $\ones(\k,\tuple{\sqnUE, \r})$ & \Cref{sec:spec:protos} & \spec{TS\,133.102}{6.3.3} \\
    \m{AKS}   & $\fives(\k,\r)$ & \Cref{sec:spec:protos} & \spec{TS\,133.102}{6.3.3} \\
    \m{AUTS} & $\tuple{\sqnUE \xor \m{AKS}, \m{MACS}}$ & \Cref{sec:spec:protos} & \spec{TS\,133.102}{6.3.3} \\
    \hline
    $\kausf$ & $\KDF(\tuple{\m{CK},\m{IK}},
               \tuple{\SNname,\sqn\XOR\m{AK}})$ & \Cref{sec:spec:protos} & \specTS{A.2}\\
    $\kseaf$ & $\KDF(\kausf,\SNname)$ & \Cref{sec:spec:protos} & \specTS{A.6}\\
    $\KSEAF(\k,\r,\sqnHN,\SNname)$ & $\kseaf$ & \Cref{sec:spec:protos} & None \\
    \hline
  \end{tabular}
  \vspace{10pt}
  \caption{Notations and Messages}
  \label{tab:notations}
\end{table*}

\section{The EAP-AKA' Protocol}
\label{ap:EAP-AKAp}
We depict the core flow of the EAP-AKA' protocol in \Cref{fig:EAP-AKAp}. We omit
the MAC failure and Re-synchronization failure phases that are the same as for 5G-AKA (see \Cref{sec:proto:5G-AKA} and \Cref{fig:5GAKA}).
We also omit the message headers specific to the EAP framework such as \cst{EAP Request} and \cst{EAP Success}.
The key derivation is a bit different compared to 5G AKA.
$\kseaf$ is derived from $\kausf$ exactly as in 5G AKA:
$$\kseaf=\KDF(\kausf, \SNname)$$
but $\kausf$ is derived differently:
$\kausf=\textsf{MK}[1152...1407]$
(we write $[x..y]$ for the substring from bit $x$ to $y$)
where the master key $\textsf{MK}$ is:
$$\KDF(\tuple{\KDF(IK,\SNname),\KDF(CK,\SNname)},
\tuple{\cst{EAP-AKA'},\supi}).$$
Therefore,
$$
\begin{array}[]{rcl}
  \kseaf &=& \KSEAF(\k,\r,\sqnHN\SNname, \supi) \\
         &= &\KDF(\KDF(\tuple{\KDF(IK,\SNname),\KDF(CK,\SNname)},\\
         &&\phantom{\KDF(\KDF(} \tuple{\cst{EAP-AKA'},\supi})[512..767],\\
         &&\phantom{\KDF(} \SNname).\\
\end{array}
$$
The messages \textsf{AT\_MAC} are MAC messages over the other sub-messages as part of the same message. The key in use is
$\kauth = \textsf{MK}[1152...1407]$.


Conceptually, the main difference of EAP-AKA' compared to 5G AKA are as follows:
\begin{itemize}
\item the challenge \m{xRES} does not directly bind the SN's identity $\SNname$. However, since the challenge is MACed (with the session key $\kauth$) together with $\SNname$, both are {\it de facto} bound together.
\item \SN serves as a pass-through until the authentication is considered successful by the \HN. Only at this time \SN obtains $\supi$ and $\kseaf$ from the \HN, while it obtains $\kseaf$ already in the first message in 5G AKA.
\end{itemize}

\definecolor{gris}{gray}{0.85}
\begin{figure*}[t]
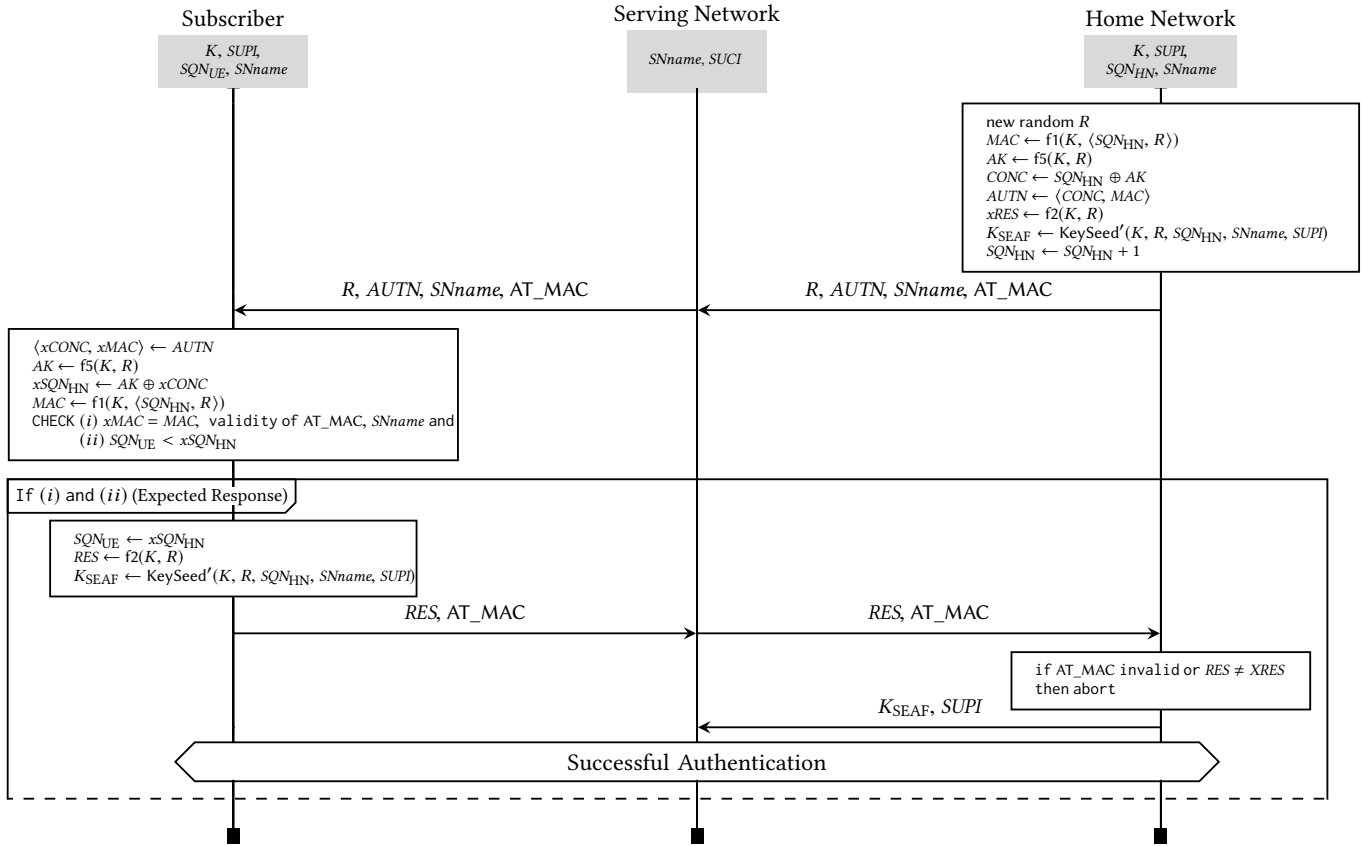

  \centering
  \setmsckeyword{}
  \drawframe{no}    
  \begin{msc}[
    /msc/title top distance=0cm,
    /msc/first level height=.2cm,
    /msc/last level height=0.4cm,
    /msc/head height=0cm,
    /msc/instance width=0cm,
    /msc/head top distance=0.5cm,
    /msc/foot distance=-0.0cm,
    /msc/instance width=0cm,
    /msc/condition height=0.2cm
    ]
    {}
    \setlength{\instwidth}{0\mscunit} 
    \setlength{\instdist}{6cm}  

   \declinst{UE}{              
      \begin{tabular}[c]{c}
        Subscriber \\ 
        \colorbox{gris}{\scriptsize{$\begin{array}{c}\k, \supi, \\ \sqn_{\UE}, \SNname \end{array}$}} 
      \end{tabular}
    }{}
    \declinst{SN}{              
      \begin{tabular}[c]{c}
        Serving Network \\
        \colorbox{gris}{\scriptsize{$\begin{array}{c}\raisebox{-5pt}{\SNname, \suci}\\ \phantom{k, \supi}\end{array}$}}
      \end{tabular}
    }{}
    \declinst{HN}{              
      \begin{tabular}[c]{c}
        Home Network \\ 
        \colorbox{gris}{\scriptsize{$\begin{array}{c}\k, \supi, \\ \sqn_{\HN}, \SNname \end{array}$}} 
      \end{tabular}
    }{} 

    \action*{\parbox{5cm}{\scriptsize{$
          \begin{array}[c]{l}
            \mathsf{new~random}\ \r \\
            \MAC\leftarrow \one(\k,\tuple{\sqnHN, \r}) \\
            \AK\leftarrow \five(\k,\r) \\
            \CONC\leftarrow \sqnHN \xor \AK \\
            \AUTN\leftarrow  \tuple{\CONC, \MAC} \\
            \m{xRES}\leftarrow \two(\k,\r) \\
            \kseaf \leftarrow \KSEAFp(\k, \r, \sqnHN, \SNname, \supi) \\
            \sqnHN \leftarrow \sqnHN + 1
          \end{array}
          $}}}{HN}
    \nextlevel[5.5]
    \mess{\messLabel{$\r,\AUTN, \SNname, \textsf{AT\_MAC}$}}{HN}{SN}
    \mess{\messLabel{$\r,\AUTN, \SNname, \textsf{AT\_MAC}$}}{SN}{UE}
    \nextlevel[0.5]

    \action*{\parbox{5.7cm}{\scriptsize{$
          \begin{array}[c]{l}
            \tuple{\m{xCONC}, \m{xMAC}} \leftarrow \AUTN\\
            \AK\leftarrow \five(\k,\r) \\
            \m{x}\sqnHN \leftarrow \AK\xor \m{xCONC}\\
            \MAC\leftarrow  \one(\k, \tuple{\sqnHN, \r}) \\
            \mathtt{CHECK}\; (i)\; \m{xMAC}=\MAC,\;\mathtt{validity\ of}\; \textsf{AT\_MAC},\SNname\;\mathtt{and}\;\\
            \phantom{\mathtt{CHECK}}\;  (ii)\; \sqnUE < \m{x}\sqnHN  
            \\
          \end{array}
          $}}}{UE}
    \nextlevel[4]

    \inlinestart[3.0cm][63pt]{thenThen}{\footnotesize $\mathtt{If}\;(i)\;\mathtt{and}\;(ii)$
      {\setlength{\fboxsep}{0pt}\colorbox{white!30}{(Expected Response)}}}{UE}{HN}
    \nextlevel[1.1]
    \action*{\parbox{4.6cm}{\scriptsize{$
          \begin{array}[c]{l}
            \sqnUE \leftarrow \m{x}\sqnHN\\
            \m{RES}\leftarrow \two(\k,\r) \\
            \kseaf \leftarrow \KSEAFp(\k,\r,\sqnHN,\SNname,\supi) \\
          \end{array}
          $}}}{UE}
    \nextlevel[3]
    \mess{\small{$\m{RES}, \textsf{AT\_MAC}$}}{UE}{SN}
    \mess{\small{$\m{RES}, \textsf{AT\_MAC}$}}{SN}{HN}
    \nextlevel[0.5]
    \action*{\parbox{3.7cm}{\scriptsize{$
          \begin{array}[c]{l}
            \mathtt{if}\  \textsf{AT\_MAC}\ \mathtt{invalid\ or}\ \m{RES} \neq \m{XRES}\\
            \mathtt{then}\ \mathtt{abort}
          \end{array}
          $}}}{HN}   
    \nextlevel[2]
    \mess{\small{$\kseaf,\supi$}}{HN}{SN}
    \nextlevel[0.4]
    \condition{Successful Authentication}{UE,SN,HN}
    \nextlevel[1.5]
    \inlineend*{thenThen}



  \end{msc}
  \vspace{-10pt}
  \caption{The EAP-AKA' protocol (continuing \Cref{fig:initAuth}). \textsf{AT\_MAC} denotes a MAC over the other messages with key $\k$.}
  \label{fig:EAP-AKAp}
\end{figure*}

\section{Lowe's Taxonomy and Tamarin Modeling}
\label{ap:taxonomy}

Lowe's Taxonomy~\cite{lowe-taxonomy} notably defines
\prop{aliveness, recent aliveness, weak agreement, non-injective agreement},
and \prop{injective agreement}.
After an introductory example showing how properties are typically modelled in \tamarin, we show
how aliveness and non-injective agreement properties are modelled. The process for the other properties
is similar.

\subsection{Introductory Example: Secrecy}
As an introductory example, let us see how secrecy properties are modeled in \tamarin.
For instance, we model the property that the $\supi$ of subscribers is never revealed to the attacker.
Formally, such a property is formalized in \tamarin using the formula defined below,
where facts $\factStyle{Claim\_Secret}(a,k)$ are produced for each rule of agent $a$ (some subscriber or some \HN)
who accesses or stores the identifier $\supi$. Note that $\factStyle{K}(t)$ denotes the fact that $t$ is in the attacker's Knowledge.
\newcommand{\timeS}{\#}
\begin{definition}
  {\em Secrecy} is modeled via the following formula:\\[5pt]
  \null\hfill$
  \forall a\ t\ \timeS i.\ \factStyle{Claim\_secret}(a, t)@i \Rightarrow \lnot(\exists \timeS j.\ \factStyle{K}(t)@j).
  $\hfill\null
  \label{def:prop:secrecy}
\end{definition}

\subsection{Aliveness}
Lowe defines \prop{aliveness} as follows (excerpt from~\cite{lowe-taxonomy}):
\begin{quotation}
  We say that a protocol guarantees to an initiator A aliveness of another agent B if,
  whenever A (acting as initiator) completes a run of the protocol,
  apparently with responder B, then B has previously been running the protocol.
\end{quotation}

Let us see how this property is mathematically modelled in \tamarin.
We assume that the \tamarin model is equipped with facts
$\factStyle{Claim\_commit}(a, b, \langle A , B \rangle)$
(i.e., an agent $a$ of role $A$ claims it has established aliveness of $b$ whose role is $B$) and
$\factStyle{Claim\_running}(b, B)$ 
(i.e., an agent $b$ of role $B$ claims it has run the protocol).

\begin{definition}
  {\em Aliveness} of a role $A$ towards a role $B$ is modeled via the following formula:
  $$  \forall a\ b\ \timeS i.\ \factStyle{Claim\_commit}(a, b, \langle A , B \rangle)@i \Rightarrow$$
\\[-20pt]  \null\hfill
  $\exists \timeS j.\ \factStyle{Claim\_running}(b, B)@j.$\null
  \label{def:prop:Aliveness}
\end{definition}
Note that we do not restrict the timestamp $\timeS j$ to be before the timestamp $\timeS i$
(\eg with a constraint $\timeS j \lessdot \timeS i$)
since the trace
semantics of \tamarin already accounts for this constraint. More precisely, if, for some execution, there was 
a fact $\factStyle{Claim\_running}(b, B)@j$ but only after
the fact $\factStyle{Claim\_commit}(a, b, \langle A , B \rangle)@i$
(\ie $\timeS i \lessdot \timeS j$), then it would suffice, for falsifying the property, to consider a prefix of the considered execution
that contains
$\factStyle{Claim\_commit}(a, b, \langle A , B \rangle)@i$
but not
$\factStyle{Claim\_running}(b, B)@j$.

\subsection{Non-injective Agreement}
Another example is the \prop{non-injective agreement} property, defined as follows
(excerpt from~\cite{lowe-taxonomy}):
\begin{quotation}
  We say that a protocol guarantees to an initiator A non-injective agreement
  with a responder B on a set of data items $t_s$ (where $t_s$
  is a set of free variables appearing in the protocol description)
  if, whenever A (acting as initiator) completes a run of
  the protocol, apparently with responder B, then B has previously
  been running the protocol, apparently with A, and
  was acting as responder in his run, and the two agents
  agreed on the data values corresponding to all the variables
  in $t_s$.
\end{quotation}

We assume that the \tamarin model is equipped with facts
$\factStyle{Claim\_commit\_agr}(a, b, \langle A , B , t_s\rangle)$
(i.e., an agent $a$ of role $A$ claims it has established agreement on data $t$ with $b$ whose role is $B$) and
\linebreak[4]
$\factStyle{Claim\_running\_agr}(b, a, \langle A , B , t_s\rangle)$ 
(i.e., an agent $b$ of role $B$ claims it tries to establish agreement on data $t$ with $a$ whose role is $A$).
The above property is modelled as follows.
\begin{definition}
  {\em Non-injective agreement} on data $t$ of a role $A$ towards a role $B$ is modeled via the following formula:
  $$
  \forall a\ b\ t\ \timeS i.\ \factStyle{Claim\_commit\_agr}(a, b, \langle A , B , t_s\rangle)@i
  \Rightarrow$$
\\[-20pt]  \null\hfill
  $  (\exists \timeS j.\ \factStyle{Claim\_running\_agr}(b, a, \langle A , B , t_s\rangle)@j).
  $\null
  \label{def:prop:NIagreement}
\end{definition}





\section{Security Assumptions and Goals}
\label{ap:prop}

\begin{center}
  \it
  This section extends \Cref{sec:spec:prop}.
\end{center}
This section is dedicated to our interpretation of security assumptions
and goals that are relevant to authentication methods in 5G as precise formal statements.
We shall support our interpretation by relevant excerpts from Technical
Specification (TS) documents and Technical Requirement (TR) documents
issued by 3GPP. Note that we may cite documents specifying aspects
of earlier generations (3G and 4G) when relevant.

\subsection{Security Assumptions and Threat Model}
\label{ap:prop:assumptions}
\subsubsection{Assumptions on Channels}
\paragraph{Channel \SN-\HN}
As part of the E2E core network,
the channel between the \SN and \HN is supposed to provide
confidentiality,
integrity,
authenticity, and,
replay protection. Those assumptions are explicitly specified:
\quoteTS{5.9.3}{
  {\bf Requirements for E2E core network interconnection security:}
  \begin{itemize}
  \item
    The solution shall provide confidentiality and/or integrity end-to-end between source and destination network for
    specific message elements identified in the present document. For this requirement to be fulfilled, the SEPP - cf
    [2], clause 6.2.17 shall be present at the edge of the source and destination networks dedicated to handling e2e
    Core Network Interconnection Security. The confidentiality and/or integrity for the message elements is provided
    between two SEPPs of the source and destination PLMN-.
  \item The destination network shall be able to determine the authenticity of the source network that sent the specific
    message elements protected according to the preceding bullet. For this requirement to be fulfilled, it shall suffice
    that a SEPP in the destination network that is dedicated to handling e2e Core Network Interconnection Security
    can determine the authenticity of the source network.
  \item The solution shall cover prevention of replay attacks.
  \end{itemize}
}

\paragraph{Channel Subscribers-\SNs}
The channel between the subscribers and \SNs, on the radio physical layer, is subject to eavesdropping (by passive attackers)
or manipulations, interception, and injection of messages (by an active attacker).
A {\em passive attacker} listens to signaling messages (\ie messages sent on the radio physical layer) on specific bandwidths
and can therefore easily eavesdrop on all messages exchanged in its vicinity.
An {\em active attacker} sets up a fake base station to receive and send signaling messages; \eg to impersonate \SNs.
While no 5G-specific hardware is publicly available yet, we recall how easily an attacker can set-up fake base stations in 4G
using open-source and freely available software and hardware~\cite{ravi-NDSS16,Golde13}.
From now on, we shall consider active attackers, except when explicitly
stated otherwise.

\subsubsection{Assumptions on Cryptographic Primitives}
According to\linebreak[4]
\spec{TS\,33.102}{3.2,6.3.2},
the functions $\one,\ones,\two$ are message authentication functions
while $\three,\four,\five,\fives$ are key derivation functions.
To the best of our knowledge, there is no standardized, explicit security requirements for these functions.
One could infer from the informal presentation \spec{TS\,33.102}{3.2} that the former are integrity protected
and the latter are integrity and confidentiality protected.
However, since $\one$ and $\ones$ are used to MAC sensitive pieces of data such as $\sqn$ (see the Section dedicated
to privacy in \Cref{ap:prop:privacy}), it is our understanding that they should additionally
preserve the confidentiality of their inputs.

Therefore, we assume $\one,\ones,\three,\four,\five,\fives$ are integrity and confidentiality protected
while $\two$ is integrity protected. We also stress that $\one$ and $\ones$ are underspecified.

\subsubsection{Assumptions on Parties}
We consider compromised scenarios in order to provide stronger and more fine-grained guarantees.
Our analyses will be parametrized by those compromised scenarios; in the worst case, a property will hold
only when the attacker cannot compromise any agent.
First, we consider an attacker who can compromise certain \SNs.
This means that the attacker gets access to an authenticated channel between the compromised \SN and
\HNs, which he can use to eavesdrop on and inject messages.
This is a reasonable assumption in 5G, where authentication methods should provide security guarantees even in presence
of genuine but malicious \SNs. In such situations, the \HNs may cooperate with such \SNs to authenticate some subscriber.
In practice, this may happen in roaming situations.
The following excerpt shows that in 5G, this is a threat model that should be considered (home refers to \HN and visited network refers to \SN):
\quoteTS{6.1.4.1}{
  {\bf Increased home control:}
  The authentication and key agreement protocols mandated to provide increased home control [compared to previous generations].
  The feature of increased home control is useful in preventing certain types of fraud, e.g. fraudulent
  Nudm\_UECM\_Registration Request for registering the subscriber's serving AMF in UDM that are not actually present
  in the visited network.
}
Furthermore, we consider that the attacker may have genuine \USIMs and compromised \USIMs under its control.
For those compromised subscribers, the attacker can access all secret values stored in the \USIMs; \ie $\supi$, $\k$, and
$\sqn$.
Finally, the attacker can access all long-term secrets $\k,\skHN$, and $\supi$ from compromised \HNs.

\subsubsection{Assumptions on Data Protection}

\paragraph{Subscriber credentials}
The subscriber credentials, notably the key $\k$ and the identifier $\supi$, shared between subscribers and \HNs, should be initially secret (provided they belong to uncompromised agents):
\quoteTS{3.1}{
  {\bf Subscription credential(s)}: set of values in the USIM and the ARPF, consisting of at least the long-term key(s) and the subscription identifier SUPI, used to uniquely identify a subscription and to mutually authenticate the UE and 5G core network.
}
\quoteTS{5.2.4}{
  The following requirements apply for the {\bf storage and processing of the subscription credentials} used to access the 5G
  network:
  \begin{itemize}
  \item The subscription credential(s) shall be integrity protected within the UE using a tamper resistant secure hardware
    component.
  \item The long-term key(s) of the subscription credential(s) (i.e. K) shall be confidentiality protected within the UE
    using a tamper resistant secure hardware component.
  \item The long-term key(s) of the subscription credential(s) shall never be available in the clear outside of the tamper
    resistant secure hardware component.
  \item The authentication algorithm(s) that make use of the subscription credentials shall always be executed within the
    tamper resistant secure hardware component.
  \item It shall be possible to perform a security evaluation / assessment according to the respective security
    requirements of the tamper resistant secure hardware component.
  \end{itemize}

  NOTE: The security assessement scheme used for the security evaluation of the tamper resistant secure hardware
  component is outside the scope of 3GPP specifications.
}

\paragraph{Sequence Number}
The sequence number $\sqn$ is a 48-bit counter (a 43-bits counter in some situations, see \spec{TS\,33.102}{C.3.2}), therefore guessable with a very low probability.
We consider a reasonable threat model where the value of $\sqn$ is unknown
to the attacker when the attack starts, but the attacker knows how it is
incremented during the attack.
This corresponds to an attacker who
(i) can monitor the activity of targeted subscribers in its vicinity during the attack but,
(ii) cannot guess the initial value of $\sqn$,
(iii) nor he can monitor targeted subscribers all the time (\ie from the very first use of the \USIM up to the attack time).
\quote{TS\,33.102}{6.3.7}{(3G)
Sequence numbers (SQN) shall have a length of 48 bits.
}

\paragraph{Other data}
While not explicitly stated in the specification, we shall assume that the private
asymmetric key $\skHN$ is initially private.


\subsection{Security Requirements}
We now extract and interpret from the 5G documents the security goals the authentication method 5G-AKA should achieve according to the 5G standard.

\subsubsection{Authentication Properties}
5G specifications make semi-formal claims about authentication properties at different places in the documents. We have identified
relevant claims and translate them into formal security goals, indicated in \prop{purple and cursive text}.
When doing this, we rely on Lowe's taxonomy of authentication properties~\cite{lowe-taxonomy}.
The first benefit is that the Lowe's taxonomy provides precise properties that are now well established and understood, which can very often
clarify an ambiguity~\cite{basin2015improving}.
The second benefit is that there exists a formal relation between the Lowe's taxononmy and mathematical definitions of security properties
that can be directly modeled in \tamarin~\cite{tamarin-manual}.
We give an overview of this taxonomy and its relation with mathematical formulations in \Cref{ap:taxonomy}.
Intuitively, it specifies, from an agent A's point of view, four levels of authentication between two agents A and B: (i) aliveness, which only ensures that B has been running the protocol previously, but not necessarily with A; (ii) weak agreement, which ensures that B has previously been running the protocol with A, but not necessarily with the same data; (iii) non-injective agreement, which ensures that B has been running the protocol with A and both agree on the data; and (iv) injective agreement, which additionally ensures that for each run of the protocol of an agent there is a unique matching run of the other agent, which prevents replay attacks.

We start by recalling the (informal) goals of authentication methods in 5G:
\quoteTS{6.1.1.1}{
The purpose of the primary authentication and key agreement procedures is to enable mutual authentication between the
UE and the network and provide keying material that can be used between the UE and network in subsequent security
procedures. The keying material generated by the primary authentication and key agreement procedure results in an
anchor key called the KSEAF provided by the AUSF of the home network to the SEAF of the serving network.
}
As we shall see, 5G aims at providing stronger guarantees than in older generations, \eg than in 3G:
\quote{TS\,133.102}{5.1.2}{(3G)
  The following security features related to entity authentication are provided:
  \begin{itemize}
  \item user authentication: the property that the serving network corroborates the user identity of the user;
  \item network authentication: the property that the user corroborates that he is connected to a serving network that is
    authorised by the user's HE to provide him services; this includes the guarantee that this authorisation is recent.
  \end{itemize}
}

We now list the security goals in terms of authentication by pairs of entities.
Note that the specification considers some authentication properties to be {\em implicit}.
This means that the guarantee is provided only after an additional {\em key confirmation roundtrip} (w.r.t.~$\kseaf$)
between the subscribers and the \SN. We discuss and criticize this design choice in \Cref{sec:analysis}.

\paragraph{Authentication between subscribers and \HNs}
First, the subscribers must have the assurance that authentication can only be successful with \SNs authorized by their \HNs.
\quoteTS{5.1.2}{
  {\bf Serving network authorization by the home network:} Assurance shall be provided to the UE that it is connected to a
  serving network that is authorized by the home network to provide services to the UE. This authorization is `implicit' in
  the sense that it is implied by a successful authentication and key agreement run.
}
\quoteTS{ 6.1.1.3}{
  The binding to the serving network prevents one serving network from claiming to be a different serving
  network, and thus provides implicit serving network authentication to the UE.
}
Formally, a \prop{subscriber must obtain non-injective agreement on $\SNname$ with its \HN after key confirmation.}
\smallskip{}

In 5G, the trust assumptions are balanced differently than in previous standards (\eg 3G or 4G). Most notably, the level of trust
the system needs to put into \SNs has been reduced. One important property provided by 5G
is that a \SN can no longer fake authentication requests
with the \HNs for subscribers that are not attached to one of its base station:
\quoteTS{6.1.4.1}{
  {\bf Increased home control:}
  The authentication and key agreement protocols mandated to provide increased home control [compared to previous generations].
  The feature of increased home control is useful in preventing certain types of fraud, e.g. fraudulent
  Nudm\_UECM\_Registration Request for registering the subscriber's serving AMF in UDM that are not actually present
  in the visited network.
}
Formally, \prop{the \HNs obtain aliveness of its subscribers at that \SN,
which is non-injective agreement  on \SNname from the \HNs' point of view with the subscribers}.

\paragraph{Authentication between subscribers and \SNs}
As expected, the \SNs shall be able to authenticate the subscribers:
\quoteTS{5.1.2}{
  {\bf Subscription authentication}: The serving network shall authenticate the Subscription Permanent Identifier (SUPI) in
  the process of authentication and key agreement between UE and network.
}
Formally, \prop{the \SNs must obtain non-injective agreement on $\supi$ with the subscribers,
  which is weak agreement from the \SNs towards subscribers} (since the $\supi$ is the subscriber's identifier).
\smallskip{}

Conversely, the subscribers shall be able to authenticate the \SNs:
\quoteTS{5.1.2}{
  {\bf Serving network authentication}: The UE shall authenticate the serving network identifier through implicit key authentication.

  NOTE 1: The meaning of 'implicit key authentication' here is that authentication is provided through the successful
  use of keys resulting from authentication and key agreement in subsequent procedures.

  NOTE 2: The preceding requirement does not imply that the UE authenticates a particular entity, e.g. an AMF, within a serving network.
}
Formally, and because $\SNname$ is the \SN's identifier,
\prop{the subscribers must obtain weak agreement with the \SNs after key confirmation.}

\paragraph{Authentication between \SNs and \HNs}
The \SNs shall be able to authenticate the subscribers that are authorized by their corresponding \HN:
\quoteTS{5.1.2}{
  {\bf UE authorization}: The serving network shall authorize the UE through the subscription profile obtained from the home network. UE authorization is based on the authenticated SUPI.
}
Formally, \prop{the \SNs must obtain non-injective agreement on $\supi$ with the \HNs}.


\subsubsection{Confidentiality Properties}
While it is not clearly specified, it is obviously the case that 5G authentication methods should achieve
\prop{secrecy of $\kseaf$, $\k$, and $\skHN$}. We recall similar goals for 3G:
\quote{TS\,133.102}{5.1.3}{(3G)
  The following security features are provided with respect to {\bf confidentiality of data} on the network access link:
  \begin{itemize}
  \item cipher algorithm agreement: the property that the MS and the SN can securely negotiate the algorithm that
    they shall use subsequently;
  \item cipher key agreement: the property that the MS and the SN agree on a cipher key that they may use subsequently;
  \item confidentiality of user data: the property that user data cannot be overheard on the radio access interface;
  \item confidentiality of signalling data: the property that signalling data cannot be overheard on the radio access
    interface.
  \end{itemize}
}
\smallskip{}

5G should ensure that knowing the $\kseaf$ established in a certain session is insufficient to deduce a $\kseaf$ key that has been
established in a previous session or that will be established in a later session:
\quoteTS{3}{
  {\bf backward security}: The property that for an entity with knowledge of $K_n$, it is computationally infeasible to compute
  any previous $K_{n-m}$ ($m>0$) from which $K_n$ is derived.

  NOTE 5: In the context of $K_{\mathrm{gNB}}$ key derivation, backward security refers to the property that, for a gNB with
  knowledge of a $K_{\mathrm{gNB}}$, shared with a UE, it is computationally infeasible to compute any previous $K_{\mathrm{gNB}}$ that
  has been used between the same UE and a previous gNB.  
}
\quoteTS{3}{
  {\bf forward security}: The property that for an entity with knowledge of Km that is used between that entity and a second
  entity, it is computationally infeasible to predict any future $K_{m+n}$ ($n>0$) used between a third entity and the second entity.

  NOTE 6: In the context of $K_{\mathrm{gNB}}$ key derivation, forward security refers to the property that, for a gNB with
  knowledge of a $K_{\mathrm{gNB}}$, shared with a UE, it is computationally infeasible to predict any future $K_{\mathrm{gNB}}$ that will
  be used between the same UE and another gNB. More specifically, n hop forward security refers to the
  property that a gNB is unable to compute keys that will be used between a UE and another gNB to which
  the UE is connected after n or more handovers ($n=1$ or more).
}
Since we do not consider the full key hierarchy and how $K_{\mathrm{gNB}}$ can be derived from $\kseaf$, we shall consider those properties
for $\kseaf$ directly.
Formally, it should be the case that \prop{$\kseaf$ established in a given session remains confidential even when the attacker
  learns the $\kseaf$ keys established in all other sessions}.
Note that this is different from \prop{forward secrecy} and \prop{post-compromise secrecy~\cite{cohn2016post}}
which fail to hold as we shall see in \Cref{sec:analysis:res}.
\smallskip{}
  
Note that some other confidentiality properties are considered to be privacy properties (see \Cref{ap:prop:privacy}).

\subsubsection{Privacy Properties}
\label{ap:prop:privacy}
We first emphasize the importance given to privacy in 5G:
\quote{TR\,33.899}{4.1,4.2}{
  Subscription privacy deals with various aspects related to the protection of subscribers' personal
  information, e.g. identifiers, location, data, etc.
  [...]
  The security mechanisms defined in NextGen shall be able to be configured to protect subscriber's privacy.
}
\quote{TR\,33.899}{5.7.1}{
  The subscription privacy is very important area for Next Generation system as can be seen by the growing attention
  towards it, both inside and outside the 3GPP world.

  Outside the 3GPP, an alliance of mobile network operators, vendors, and universities called NGMN [9] has identified
  security and privacy as an enabler and essential value proposition of NextGen system and has presented that built-in
  privacy should be included as a design principle [10]. Similarly, a 5G PPP project called 5G-Ensure [11] has also
  identified privacy as one of the topmost priorities for the NextGen system stating that the privacy has an important
  social impact [12].
  [...]
}
\quoteTS{F.2}{
  EAP-AKA' includes optional support for identity privacy mechanism that protects the privacy against passive
  eavesdropping.
}
This important role given to privacy can be explained by numerous and critical attacks that have breached privacy (\eg with IMSI-catcher~\cite{ravi-NDSS16,Broek2015})
in previous generations; see the survey~\cite{mobileSoK17}.

We also recall that privacy was already a concern in 3G:
\quote{TS\,133.102}{5.1.1}{(3G)
  ~The following security features related to user identity confidentiality are provided:
  \begin{itemize}
  \item  {\bf user identity confidentiality}: the property that the permanent user identity (IMSI) of a user to whom a services
    is delivered cannot be eavesdropped on the radio access link;
  \item  {\bf user location confidentiality}: the property that the presence or the arrival of a user in a certain area cannot be
    determined by eavesdropping on the radio access link;
  \item  {\bf user untraceability}: the property that an intruder cannot deduce whether different services are delivered to the
    same user by eavesdropping on the radio access link.
  \end{itemize}
}
Thus, already for 3G, user identity confidentiality, anonymity, and untraceability
were security requirements. However, those properties were required against a passive attacker only
(we discuss and criticize this restriction to a passive attacker in \Cref{sec:analysis}).
Note that anonymity and untraceability (often called unlinkability) are not clearly defined. We propose formalization in \Cref{sec:formal:choices}.  

We now list more precise statements specifying how privacy should be protected in 5G.

\paragraph{Confidentiality of $\supi$}
In 5G, the $\supi$ is considered sensitive and must remain secret since it uniquely identifies users.
Indeed, an attacker who would be able to obtain this value from a subscriber would be able to
identify him, leading to classical user location attacks (\ie see \spec{TS\.133.102}{5.1.1} above),
much like IMSI-catcher attacks.
\quoteTS{5.2.5}{
  The SUPI should not be transferred in clear text over 5G RAN except routing information, e.g. Mobile Country Code
  (MCC) and Mobile Network Code (MNC).  
}
\quoteTS{6.12}{
  {\bf Subscription identifier privacy:}
  In the 5G system, the globally unique 5G subscription permanent identifier is called SUPI as defined in 3GPP TS
  23.501 [2]. The SUCI is a privacy preserving identifier containing the concealed SUPI.
  [...]
}
\quote{TS\,133.102}{5.1.1}{(3G)
{\bf User identity confidentiality} (see above).
}
Formally, \prop{the $\supi$ shall remain secret in the presence of a passive attacker}.

\paragraph{Confidentiality of $\sqn$}\
For similar reasons, $\sqn$ must remain secret. An additional reason that is not explicitly stated
is that $\sqn$ leaks the number of successful authentications the corresponding \USIM has performed
since it was manufactured, which is strongly correlated to its age and activity. This is even more
critical when the attacker learns $\sqn$ at different times.
\quote{TS\,33.102}{6.2.3}{(3G)
  Here, AK is an anonymity key used to conceal the sequence number as the latter may expose the identity and location
  of the user. The concealment of the sequence number is to protect against passive attacks only. If no concealment is
  needed then f5 $\equiv$ 0 (AK = 0).
}
\quote{TS\,133.102}{C.3.2}{(3G)
  {\bf User anonymity}: the value of SQN may allow to trace the user over longer periods. If this is a concern then SQN has to
  be concealed by an anonymity key as specified in section 6.3.
}
Formally, \prop{the $\sqn$ shall remain secret in the presence of a passive attacker}.

\paragraph{Anonymity and Untraceability}
Preventing the attacker from learning pieces of data that are identifying (\eg $\supi$, $\sqn$) is insufficient
to protect against traceability attacks, user location attacks, or even anonymity attacks (we explain why
and discuss definitions in \Cref{sec:formal:choices}).
While no formal statement is made on the necessity of ensuring untraceability or anonymity for 5G,
the following excerpts and the fact that it was required for 3G (\spec{TS\.133.102}{5.1.1}, see above),
seem to imply that those properties are relevant for 5G as well.

On untraceability (also called unlinkability): \spec{TS\,133.102}{5.1.1}\textsf{, item 2} and:
\quoteTS{C.2}{
  The reason for mentioning the non-freshness is that, normally, in order to attain {\bf unlinkability} (i.e., to
  make it infeasible for over-the-air attacker to link SUCIs together), it is necessary for newly generated
  SUCIs to be fresh. But, in case of the null-scheme, the SUCI does not conceal the SUPI. So unlinkability
  is irrelevant.
}
\quote{TR\,33.899}{5.2.3.8.2}{
  {\bf Security threats:}
  Over-use of a single UE key-pair may harm user privacy (allowing a user's actions to be linked and tracked
  across multiple domains and services).
}

On anonymity: \spec{TS\,133.102}{5.1.1}\textsf{, item 3} and:
\quote{33.849}{6.4.2}{(TR on Privacy in 3GPP, 2016)
  The UMTS authentication procedure (TS 33.102 [10]) design is an example of how to fulfil anonymity:
  \begin{enumerate}
  \item Analysis of the authentication process: identity and location of the user may be exposed.
  \item Identify an identifying attribute: sequence number may bring a risk of personal identification.
  \item Risk: The {\bf sequence number may expose the identity and thus the location of the user}.
  \item Anonymizing technique used: use Anonymity Key in the Authentication Token to conceal (blind) the sequence number.
  \end{enumerate}}
\quote{TR\,33.899}{5.1.4.14.3}{
  If there was no single NAS security termination then the unencrypted part of a signalling message would have to
  contain parameters that would allow routing to the correct NAS entity, e.g. SM entity in a network slice. This
  information about the slice may give away information on the services used. However, user identity privacy should
  prevent that an eavesdropper can associate a particular signalling message with a particular subscriber.
  Editor's Note: The above paragraph has been included for completeness. It is ffs whether leaving parameters
  unencrypted that are required for NAS-internal routing would endanger privacy.
}

Formally, it seems that 5G authentication methods are required to provide \prop{anonymity and untraceability of
  the subscribers in the presence of a passive attacker}.


\subsubsection{Other Properties}
As specified below, the established keys should never be the same twice:
\quote{TS\,133.102}{6.2.3}{(3G)
  {\bf Key reuse:}
  A wrap around of the counter SQN could lead to a repeated use of a key pair (CK, IK). This repeated key
  use could potentially be exploited by an attacker to compromise encryption or forge message
  authentication codes applied to data sent over the 3GPP-defined air interfaces.
}
This will be analyzed as part of \prop{{\em Injective} agreement properties on the established key $\kseaf$}
for different pairs of parties.
\smallskip{}

Finally, 5G specify some security goals in the context of backward compatibility with older generations.
We do not analyze those properties as they would require us to analyze the combination of the 5G authentication
protocols with the older generations authentication protocols. This is left as future work.
\quoteTS{6.1.1.3}{
  {\bf Key separation:}
  Furthermore, the anchor key provided to the serving network shall also be specific to the authentication having taken
  place between the UE and a 5G core network, i.e. the KSEAF shall be {\bf cryptographically separate} from the key KASME
  delivered from the home network to the serving network in earlier mobile network generations.
}
\quoteTS{5.11}{
  An attacker could attempt a {\bf bidding down attack} by making the UE and the network entities respectively believe that
  the other side does not support a security feature, even when both sides in fact support that security feature. It shall be
  ensured that a bidding down attack, in the above sense, can be prevented.
}

\balance

\end{document}